\documentclass[12pt]{article}%
\usepackage{amsmath}
\usepackage{mathtools}
\usepackage[onehalfspacing]{setspace}
\usepackage{multicol}
\usepackage{natbib}
\usepackage{amsfonts}
\usepackage{amssymb}
\usepackage{graphicx}
\usepackage{caption}
\usepackage{xcolor}
\usepackage{multirow}
\usepackage{array}
\usepackage[colorlinks=true,linkcolor=blue,citecolor=blue]{hyperref}
\usepackage{nicematrix}
\usepackage[a4paper, total={6in, 8in}]{geometry}
\usepackage{float}%

\providecommand{\U}[1]{\protect\rule{.1in}{.1in}}
\newtheorem{theorem}{Theorem}

\newtheorem{lemma}[theorem]{Lemma}

\newtheorem{proposition}[theorem]{Proposition}

\newenvironment{proof}[1][Proof]{\noindent\textbf{#1.} }{\ \rule{0.5em}{0.5em}}
\begin{document}

\title{Bribery, Secrecy, and Communication: \\ Theory and Evidence from Firms\thanks{We would like to thank Yaron Azrieli, Daniel Berkowitz, Paul
Healy, Natalia Kravtsova, Kurt Lavetti, Alejandra Lopez, Farhod Olimov, Brian Roe, Jason Tayawa, Bruce Weinberg, Matthew Weinberg  and seminar participants at the Ohio State University. Declarations of interest: none. }}
\author{Jafar M. Olimov\thanks{Address: Department of Economics, The Ohio State
University, 435 Arps Hall, 1945 N High Street, Columbus, Ohio, 43210; email:
olimov.1@osu.edu}\\The Ohio State University}
\date{November 2025}
\maketitle

\begin{abstract}

This paper studies if firms pay different types of bribes, and if corrupt bureaucrats have perfect information about resources of bribe-paying firms. We construct a model of corruption that allows for multiple informational scenarios in a single market for bribes and empirically test these scenarios on the original dataset of 429 firms operating in Tajikistan. The results indicate that firms simultaneously make voluntary and involuntary bribe payments, firms hide resources from corrupt bureaucrats to reduce involuntary bribe payments, and bureaucrats who receive voluntary bribe payments do not share bribery-relevant information with other bureaucrats. 

\textbf{Keywords: }Bribes, Red tape, Firm behavior, Corruption

\textbf{JEL Codes:} D22, D73, D82, K42, O12

\end{abstract}

\newpage

\section{Introduction}\label{sec:intro}

The prominent approach in the literature on corruption is to view the market for bribes as centralized, with a single bureaucrat receiving bribes of one type in return for a public service that should be offered bribe-free. The literature offers two mutually exclusive perspectives on how to look at bribery. Under the ``grease the wheels" bribery hypothesis, firms voluntarily make bribe payments to a corrupt bureaucrat to avoid burdensome bureaucratic barriers (\cite{Huntington2006}, \cite{Leff1964}) or to obtain limited government resources (\cite{Vial2010}, \cite{Lui1985}). Under this hypothesis, bribery is viewed as beneficial both socially and at the individual firm level. Under the competing
``sand in the wheels" bribery hypothesis, the corrupt bureaucrat extorts bribes from unwilling firms by threatening to subject them to costly red tape (e.g. \cite{BanerjeeMullainathanHanna2013}, \cite{Banerjee1997}). Under this hypothesis, bribery is viewed as harmful at the aggregate and individual firm levels (\cite{Uberti2022}, \cite{Paunov2016}, \cite{Mauro1995}). 

We present evidence that firms pay both types of bribes: non-extortionary (consistent with the ``grease" hypothesis) and extortionary (consistent with the ``sand" hypothesis). The coexistence of distinct types of bribery indicates the presence of at least two distinct groups of corrupt bureaucrats in the same market for bribes. We do not find evidence that bureaucrats engaging in distinct types of bribery share bribery-relevant information. In particular, we find that bureaucrats involved in the non-extortionary bribery do not share information about firms' financial resources with bureaucrats involved in the extortionary bribery. These results indicate that the assumption of a centralized market for bribes may not always hold, and the commonly used measures of aggregate bribes in the empirical literature possibly aggregate over multiple types of bribes.  

To study the coexistence of bribery types, we construct a theoretical model of corruption that incorporates distinct types of bribery, analyze it under different informational scenarios, and empirically test which of the scenarios is consistent with the observed behavior of firms. We study the equilibrium behavior of firms (\cite{Basu2020}).

In our model of corruption, a single firm with hidden financial resources can pay two types of bribes (as in \cite{KhalilLawareeYun2010}) and can face bureaucratic delay (as in \cite{Guriev2004} and \cite{Banerjee1997}). The firm can willingly participate in a contest to gain access to a public resource (e.g. obtain a procurement contract) by paying a non-extortionary bribe to a corrupt bureaucrat who administers the contest in the first period. The same firm faces an extortionary request for a bribe that the firm is not willing to pay (e.g. when submitting an annual tax report) from another bureaucrat in the second period. 

We consider three informational scenarios. In the ``no secrecy" scenario, the bureaucrat in the second period perfectly observes the firm's type and entry into the contest in the first period. In the ``secrecy with communication" scenario, the bureaucrat in the second period does not observe the firm's type but receives a message about the firm's entry into the contest from the bureaucrat in the first period. In the ``secrecy without communication" scenario, the bureaucrat in the second period neither observes the firm's type nor receives a message about the firm's entry into the contest from the bureaucrat in the first period. 

Based on the equilibrium conditions of the model of corruption (Propositions \ref{prop:eq_bribe_NS}, \ref{prop:eq_bribe_SC}, and \ref{prop:eq_bribe_SwC}), we formulate testable specifications (\ref{eq:EM1}, \ref{eq:EM11}) to answer three questions: (1) whether firms pay both non-extortionary and extortionary bribes (multiplicity of bribery types), (2) whether firms hide at least a portion of their financial resources from corrupt bureaucrats (secrecy of firms' resources), and (3) whether bureaucrats receiving voluntarily paid (non-extortionary) bribes share bribery-relevant information with bureaucrats that extract involuntarily paid (extortionary) bribes (communication in the market for bribes). 

The empirical analysis in our paper is related to firm-level studies of corruption (e.g. \cite{FismanGurievIoramashviliPlekhanov2023},
\cite{ColonnelliPrem2021},
\cite{BaiJayachandranMaleskyOlken2017},
\cite{FreundHallwardDriermeierRijkers2016},
\cite{SequeiraDjankov2014},
\cite{RandTarp2012}). In contrast to existing empirical studies, we do not make causal arguments about the determinants or consequences of corruption. Instead, we quantify frictions that set the scope for different types of bribery. In this regard, this paper is more closely related to firm-level studies that directly test mechanisms that model bribery. These studies include \cite{Svensson2003}, who tests the bargaining hypothesis of bribery and finds that firms with smaller observable profit levels and larger outside options measured by market value of capital are able to
negotiate lower bribes, and \cite{OlkenBarron2009}, who find evidence that corrupt officials exploit their monopoly power to extract larger bribes by using two-part tariffs and third-degree price discrimination.

To carry out empirical tests, we use the original data from a nationally representative survey of $429$ firms operating in 2012-2014 in Tajikistan, where corruption is a common phenomenon\footnote{According to Transparency International, in 2015 Tajikistan was ranked 136th out of 168 countries in terms of corruption perception (source: 
\url{ https://www.transparency.org/en/cpi/2015/index/tjk}). According to the Ease of Doing Business index in 2015, Tajikistan was ranked 166th out of 189 countries in terms of the friendliness of the regulatory environment (source: World Bank (2014) Doing Business 2015: Going Beyond Efficiency. Washington, DC: World Bank Group).} (see Tables \ref{table:profit} and \ref{table:data_summary} for data summaries). When collecting data, we adopt the data collection methodology of the World Bank's Enterprise Surveys
(WBES), which collect data on business practices of firms around the world\footnote{See \url{https://www.enterprisesurveys.org/en/methodology} for
description of the data collection procedure in WBES.}. We follow the WBES's guidelines when collecting data on sensitive information such as bribe payments, unreported sales, and overstated costs. 

Our first contribution is to show that both types of bribery, non-extortionary and extortionary, coexist in the same market for bribes. We rely on distinct relationships between bribes and paperwork processing delays during and outside of bureaucratic disputes to identify different types of bribery. The coexistence of different types of bribery is the evidence for both the ``grease" and the ``sand" hypotheses holding at the same time and is in contrast to the view in the literature that the two bribery hypotheses are mutually exclusive (\cite{Martins2020},
\cite{FreundHallwardDriermeierRijkers2016}, \cite{MeonSekkat2005}, \cite{KaufmannWei1999}). The coexistence of different types of bribery is consistent with the findings in \cite{Gauthier2021}, who differentiate between demand-side and supply-side bribery and find empirical evidence that bribery in the tax administration is distinct from bribery in the public procurement. In contrast to \cite{Gauthier2021}, who assume that a single firm can engage only in one type of bribery, we assume that a single firm can simultaneously engage in multiple types of bribery, which allows us to study the informational aspect of bribery. 

Since we use the same data collection methodology as in WBES, our results indicate that the commonly used measure of bribes in WBES and similar firm-level data (annual bribe as a share of total sales) is an aggregation of bribe transfers with distinct payment mechanisms and distinct implications. Consequently, our results point out the need to account for heterogeneity in bribery types that may be present in aggregate bribe measures in the literature.

Our second contribution concerns the question of observability of resources of bribe-paying firms.  We find that more than 60 percent of firms in our data mis-report profits in tax statements by understating sales and by overstating operating costs (see Table \ref{table:profit} and related discussion). The misreporting of profits may occur for multiple reasons, including tax evasion (see \cite{Slemrod2002} for the review), complexity of tax procedures (\cite{DeNeve2021}), lax enforcement of tax regulations (\cite{Rubolino2023}), or under-developed financial system (\cite{Guo2020}). We offer one more reason why firms mis-report profits, namely because firms attempt to reduce bribe demands from corrupt government officials. 
Specifically, we find a positive statistically significant correlation between aggregate bribes and fractions of profits reported in tax statements, and no such relationship between bribes and fractions of profits unreported in tax statements for firms facing extortionary bribe demands (see Table \ref{table:coefs}). This result highlights the secrecy aspect of bribery, as indicated in \cite{ShleiferVishny1998}.

Our third contribution is to show the lack of communication about bribery-relevant characteristics of firms among bureaucrats involved in distinct types of bribery. This result is particularly important in the context of extortionary bribery with imperfect information. In the absence of information about financial resources of firms, corrupt bureaucrats demanding extortionary bribes can obtain this information from other corrupt bureaucrats. When distinct groups of corrupt bureaucrats do not communicate, the uninformed corrupt bureaucrats making extortionary bribe demands have to rely on a screening mechanism to sort out firms with more financial resources, for example, by subjecting all firms to differentiated levels of costly bureaucratic delay \cite{Banerjee1997}. While we find evidence of the use of bureaucratic delay as a screening device, we find no evidence of communication among bureaucrats involved in distinct types of bribery. 

Our results on heterogeneity of bribery types, unobservability of bribery-relevant firm characteristics, and the lack of communication among distinct groups of corrupt bureaucrats indicate that the market for bribes in this paper is more appropriately described as decentralized. Consequently, commonly made assumptions about homogeneity of bribery types and centralized bribery may not hold in arbitrary markets.

The paper is organized as follows. In Section \ref{sec:market}, we describe the market for bribes using data from firms in Tajikistan. In Section \ref{sec:model}, we present the model of corruption and derive equilibrium conditions to obtain empirical specification in Section \ref{sec:identification}. In Section \ref{sec:estimation}, we discuss the data, estimation approach, and present estimation results. We discuss results and provide policy recommendations in Section \ref{sec:discussion} and conclude in Section \ref{sec:conclusion}.

\section{The market for bribes}\label{sec:market}

We define the market for bribes as a market of all government services for
which firms either transfer money to bureaucrats or for which bureaucrats
demand money from firms on top of what is required by law, in an economy in
fiscal year. The demand side represents all firms that have at least tax
registration and excludes firms that operate completely informally (without
tax registration) since such firms may not need to interact with bureaucrats. The supply side includes all bureaucrats that either receive voluntarily made payments or demand money for their services on top of what is required by law.

\subsection{Bribes}\label{subsec:bribes}
The literature on corruption has long recognized differences in the types of bribery.
The discussion of distinct bribe types originates in the law literature, and it does not overlap with the ``grease vs. sand" bribery debate. In the law
literature, \cite{Ayres1997} defines extortion as a conspiracy initiated by
a judge to achieve a less than fair outcome for a defendant, and bribery as a
conspiracy initiated by the defendant to achieve a more than fair outcome for
himself. Both types of conspiracies involve illegal monetary transfers from
the defendant to the judge. The key distinction lies in the nature of the
relationship between the parties: the defendant transfers money to the judge voluntarily in the case of bribery and involuntarily in the case of extortion. 
\cite{KhalilLawareeYun2010} are the first to introduce this conceptual
difference into the economics literature by
characterizing bribery as a ``cooperative" transaction that involves the pursuit of a ``common objective" by the firm and the corrupt bureaucrat, and extortion as an ``antagonistic" transaction that benefits the corrupt bureaucrat at the expense of the firm. The former (cooperative) type is called \textit{collusive} bribery in the experimental economics literature (see \cite{Abbink2002}) and is consistent with \textit{bribery without theft} in the terminology of \cite{ShleiferVishny1993}. The latter (antagonistic) type is called \textit{harassment} bribery (see \cite{Basu2011}, \cite{Abbink2014}) in the experimental economics literature and is consistent with \textit{bribery with theft} in the terminology of \cite{ShleiferVishny1993}. 

Although the literature on distinct types of bribery and the ``grease vs. sand"
bribery debate do not overlap, there is no inconsistency between these concepts. Under the
``grease the wheels" hypothesis, a firm willingly transfers money to a corrupt
bureaucrat to avoid bureaucratic obstacles or to gain access to public
resources. This type of monetary transfer is consistent with bribery in the terminology of \cite{Ayres1997}. Under the ``sand in the wheels" hypothesis, a corrupt bureaucrat demands a bribe that
a firm is not willing to pay, and to incentivize the firm, the corrupt
bureaucrat uses artificial barriers such as red tape. This bribery mechanism
is consistent with the definition of extortion by \cite{Ayres1997}. 

We follow the terminology of \cite{Ayres1997} and call bribes that firms involuntarily pay to corrupt bureaucrats as \textit{extortionary}. We assume that payment of these bribes is consistent
with the ``sand in the wheels" hypothesis. We call bribes that firms
voluntarily pay to corrupt bureaucrats as \textit{non-extortionary}, and
assume that payment of these bribes is consistent with the ``grease the wheels"
bribery hypothesis.

The third type of informal transfers that are usually classified as bribes are the customary
transfers such as gifts or donations that bureaucrats receive from the
public, because transfers of such gifts are socially accepted or because they
are a cultural norm (see \cite{Malesky2020}, \cite{Dong2012}, \cite{FismanMiguel2007}). Note that customary transfers can not be explained by the
non-extortionary bribery mechanism, because firms making such transfers do not
expect any favors in return. Further, the customary transfers cannot be
explained by the extortionary bribery mechanism, because these gifts or
donations are not explicitly solicited by bureaucrats.

Despite the conceptual differences between distinct types of bribes, it is
challenging to collect data on separate bribe payments for two main 
reasons. Firstly, since different bribe types are often paid
simultaneously and there is no uniformly accepted codification of bribery,
bribe-payers fail to determine whether a bribe is paid voluntarily or
involuntarily, especially, if a bribe is a part of a larger transaction.
Secondly, subjects of firm-level studies of corruption often refuse to report
data on separate bribe payments either because they do not keep track of
separate bribe payments or because they fear that reports of separate bribe
payments may reveal information about their relationship with particular
corrupt bureaucrats. For these reasons, we follow the established approach in
the literature and collect bribery data only in the aggregate form (bribe as a share of total sales).

\subsection{Bureaucrats}
In a completely centralized market for bribes, the supply side is represented by a single bureaucrat, who sets a bribe payment for each government service for each firm with tax registration. 
In this type of market, even with multiple bureaucrats receiving or
demanding bribes, a single bureaucratic entity aggregates information about
identities and bribe payments of all bribe-payers. In particular, in a
centralized market, the single bureaucratic entity can condition later bribe
demands on prior bribe payments for each firm paying multiple bribes.

In a decentralized market, the supply side is represented by multiple
bureaucrats that individually receive or demand bribe payments. Typically,
each bribe is paid in return for a public service that each bureaucrat is
required to provide to the public. In this type of market, a bribe-receiving bureaucrat cannot condition his bribe demand on bribe payments made to other bureaucrats, unless bribe-receiving bureaucrats communicate or bribe transfers are public information.

Corrupt bureaucrats differ both in their reach of firms and in their ability
to extract bribes. Corrupt bureaucrats who provide services that all firms are
required to obtain by law have the most power to extract bribes and have the
widest reach. For example, a corrupt tax inspector has the most power to
extract bribes from all firms in his tax district, as all firms with tax
registration have to submit annual tax reports. Because operation without an
approved tax report is illegal, a corrupt tax inspector can successfully engage in extortionary bribery by extracting bribes even from firms that are not willing to pay bribes.  

In contrast, a corrupt bureaucrat administering a contest for allocation of a public procurement contract among private firms has narrow reach and the least power to extract bribes, since he can only target firms that have an interest in obtaining the contract. Furthermore, because firms can exit the contest at will, such bureaucrat's power to demand bribes is limited. As a result, a bribe paid to a corrupt bureaucrat administering the contest is more appropriately described as non-extortionary.

A single firm can pay multiple types of bribes due to multiple bureaucratic requirements\footnote{In our data, the average annual number of certificates that firms are required to obtain is 8.}. For example, a firm may pay a non-extortionary bribe when trying to obtain a government contract while simultaneously paying an extortionary bribe when filing a tax statement at the end of the fiscal year. In a more centralized market for bribes, where different corrupt bureaucrats share information and coordinate, the extortionary bribe paid to the tax inspector at the end of the fiscal year can be conditioned on the size of the non-extortionary bribe paid to the bureaucrat administering the public procurement contest. This may, in turn, affect the firm's incentive to participate in the contest and make the non-extortionary bribe payment. In a completely decentralized market for bribes, where all corrupt bureaucrats operate independently and do not share information, a firm making a non-extortionary bribe payment does not need to be concerned that its voluntary decision to pay bribes can affect bribe demands from other corrupt bureaucrats in a later period.

\subsection{Firms}
Firms differ in their ability to pay bribes. If bribe is a form of a tax on
firms' disposable financial resources (profit), then most profitable firms can afford to pay
largest bribes, whereas firms with non-positive profits can afford to pay
only small bribes or no bribes at all. With perfect knowledge of firms'
financial resources, corrupt bureaucrats with appropriate administrative
powers can completely extract firms' resources, for example,
by making take-it-or-leave-it offers to firms with positive profits while
ignoring firms with non-positive profits. By ignoring low-productivity
non-positive-profit firms, corrupt bureaucrats prevent such firms from exiting
to the informal sector (see \cite{Ulyssea2018})
or from completely shutting down operations. This is optimal from corrupt
bureaucrats' perspective, as non-positive-profit firms operating in the formal
sector can be targeted later, when their profits increase. 

Because firms differ in their ability to pay bribes and because in extortionary bribery corrupt bureaucrats are the ones making bribe demands, the extortionary bribery critically depends on the corrupt bureaucrats' knowledge of firms' resources. The problem of informational asymmetry in extortionary bribery is exacerbated by the unwillingness of firms to reveal information about their resources, as firms participate in this type of bribery involuntarily. In this
regard, corrupt tax officials are particularly apt to effectively exploit their administrative powers in extortionary bribery, as all firms with tax registration are obligated to file annual tax reports, and tax reports can potentially fully reveal firms' resources.

\begin{table}
\center{
\caption{Shares of firms with positive, zero, and negative
profits.}
\small{
\begin{tabular}
[c]{ccrrr}\hline\hline
Year & Data & Negative profits & \ \ \ \ \ Zero profits & Positive
profits\\\hline
2012 & Reported & 55.9\% & 36.5\% & 7.6\%\\
& Actual & 20.5\% & 10.4\% & 69.1\%\\
2013 & Reported & 55.1\% & 36.3\% & 8.6\%\\
& Actual & 17.9\% & 10.5\% & 71.6\%\\
2014 & Reported & 59.3\% & 33.1\% & 7.6\%\\
& Actual & 21.7\% & 9.3\% & 69.0\%\\\hline\hline
\end{tabular} \label{table:profit}
}
\caption*{\footnotesize{Note: The numbers are from the survey firms
operating in Tajikistan in 2012 - 2014. The reported profits are the differences
between firms' sales and costs reported to local tax offices. The actual profits are
the differences between reported sales adjusted for the amount of under-reported
sales and reported costs adjusted for the amount of over-stated costs, as
evaluated by the firms themselves.}}
}
\end{table}

\normalsize

In contrast to bureaucrats in extortionary bribery, corrupt bureaucrats
that receive non-extortionary bribes do not need to rely on the their
knowledge of firms' resources to elicit bribe offers. Since non-extortionary
bribery can be viewed as a contest among firms that voluntarily participate in bribery,
bribes of contest-winning firms are determined by the competition among contest
participants rather than by bribe demands of corrupt bureaucrats.\footnote{This view of
bribery is adopted in the literature on corruption in procurement
auctions. See, for example, \cite{Compte2005} for a discussion of how bribery
affects bidding behavior in procurement auctions.}. 

If firms anticipate extortionary bribe demands and expect non-positive-profit firms to pay small or no bribes, then positive-profit firms have an
incentive to imitate non-positive-profit firms, for example, by under-reporting revenues and
over-stating costs in tax reports. We can test this conjecture by directly
comparing distributions of firms by actual and reported profitability in our
data. In Table \ref{table:profit}, we report distributions of firms with negative, zero, and
positive profits for each year in the data. The rows with ``reported" numbers
present shares of firms with negative, zero, positive profits calculated from
revenues and costs submitted to local tax inspectors. As Table \ref{table:profit} shows, not all firms
report their profits truthfully. This is evidenced by the clustering of firms
reporting zero profit in the table. Further evidence is given in rows with ``actual"
numbers that report shares of firms by profitability when accounted for
under-reported sales and over-stated costs in tax reports.

From the perspective of corrupt bureaucrats, there are at least two types of
firms. The positive-profit firms that can afford to
pay bribes belong to one type. The non-positive-profit firms that cannot pay
bribes or can pay only small bribes belong to the other type. According to
Table \ref{table:profit}, about 70\% of firms in the data are of the former type, 30\% of the
firms are of the latter type, and more than 60\% of the positive-profit firms
pretend to belong to the non-positive-profit type. If corrupt bureaucrats rely
only on the reported tax data, then the actual type of about 90\% of
positive-profit firms is not observable, suggesting that corrupt bureaucrats
may have to rely either on the information about firms' resources from other bureaucrats or on a screening mechanism that could sort out the mimicking
positive-profit firms.

\begin{figure}
\small{
\begin{center}
\includegraphics[width=0.9\textwidth]{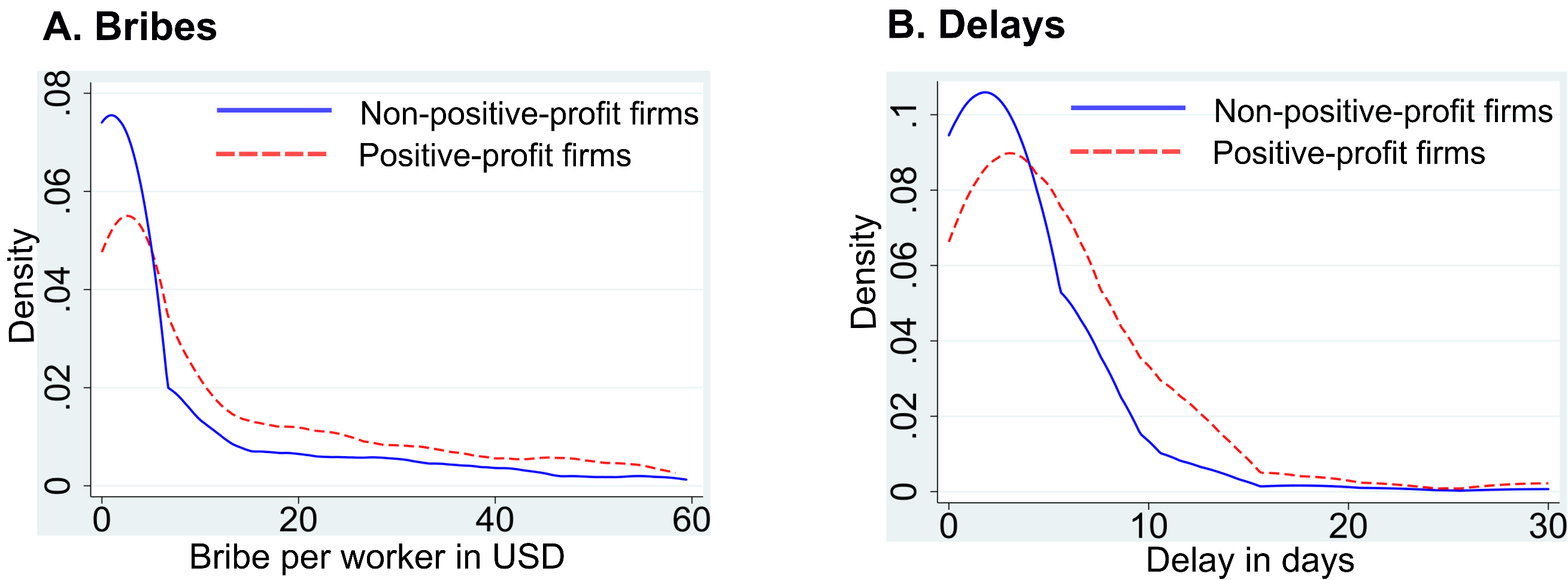}
\end{center}

 \caption{Bribes and delays (data of firms in Tajikistan in 2012 - 2014). A. Distributions of aggregate annual bribes per worker are estimated non-parametrically using Epanechnikov kernel (bandwidth=3), truncated at US $\$60$. B. Distributions of delays estimated non-parametrically using Epanechnikov kernel (bandwidth=2.5), truncated at $30$ days.}\label{fig:bribe_delay}
}
\normalsize
\end{figure}

To test whether the behavior of positive-profit firms is distinct from that of
non-positive-profit firms, we compare distributions of reported bribe payments of
positive-profit and non-positive-profit firms in Figure \ref{fig:bribe_delay}A. The behavior of
firms is clearly distinct. While 84\% of positive-profit firms report paying
bribes, only 49\% of non-positive-profit firms report doing so in the data.
Further, as Figure \ref{fig:bribe_delay}A shows, positive-profit firms tend to pay larger bribes
than non-positive-profit firms.

Because red tape is often associated with bribery (\cite{Banerjee1997}, \cite{FismanGatti2006}),
we also look at the differences in bureaucratic delays of positive-profit and
non-positive-profit firms. The pattern is similar. Only
11\% of positive-profit firms and 47\% of non-positive-profit firms report
having no bureaucratic delays. As shown in Figure \ref{fig:bribe_delay}B, bureaucratic delays are
longer for positive-profit firms than for non-positive-profit firms. These
comparisons indicate that the overwhelming majority of positive-profit firms
pay bribes (84\%) and face bureaucratic delays (89\%). This is in contrast to
the less than half of non-positive profit firms reporting paying bribes (49\%)
and facing bureaucratic delays (47\%).

\section{The model of corruption}\label{sec:model}

\subsection{Model setup}
\begin{figure}
\begin{center}
\includegraphics[width=0.9\textwidth]{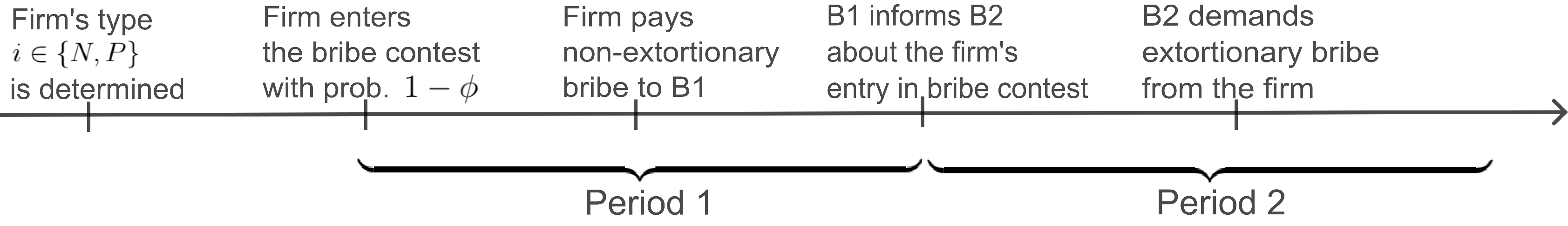}
\caption{Timing of the model of corruption.}\label{fig:timing}
\end{center}
\end{figure}

 We introduce the model of corruption with two bribery types and potentially unobservable types of a single firm. The timing of the model is in Figure \ref{fig:timing}.

\paragraph*{\textbf{Types of firm:}} We denote the firm's type by $i\in\{N,P\}$, where $i=P$ represents a firm with positive profit and $i=N$ represents a firm with non-positive profit. The firm's profit is denoted by $\Pi_i$. Bureaucrats' believe that the firm has
non-positive profit with probability $\mathbb{P}(i=N)=\theta$ and positive profit
with probability $\mathbb{P}(i=P)=1-\theta$, $\theta \in [0,1]$. We express the $P$-type firm's profit as the sum of the $N$-type firm's profit and a positive increment $\Delta\Pi$, i.e. 
$\Pi_{P}=\Pi_{N}+\Delta\Pi$. 

\paragraph*{\textbf{Types of bribes: }} The firm obtains two services: a contract $\Gamma$ from bureaucrat $B1$ in
period $1$ and a certificate from bureaucrat $B2$ in period $2$. Although the
firm can obtain the contract $\Gamma$ from $B1$ at will, it is required to
obtain the certificate from $B2$ independently of the acquisition of the contract in period $1$.  

The acquisition of both services is costly and may involve bribe payments and bureaucratic
delays. We label the bribe paid
to $B1$ as non-extortionary and the bribe paid to $B2$ as extortionary.  Since data on individual
bribe payments are not available, we focus on aggregate bribes. An aggregate bribe is the sum of non-extortionary and extortionary bribes paid by the firm
in both periods.

\paragraph*{Non-extortionary bribery in period 1:}

The firm decides whether to obtain contract $\Gamma\in\{0,1\}$, with $\Gamma=1$ for the firm obtaining the contract, and $\Gamma=0$ otherwise. Since $B1$ is corrupt, the allocation of $\Gamma$ is determined in a bribe contest, where a single firm or many firms offer bribes to $B1$ in return for the contract. The contract can be awarded to all firms offering bribes or only to a fraction of such firms. The net payoff from the contract to firms is $V$, and $V$ is known to $B1$. 

The firm makes the decision to enter the bribe contest with probability $1-\phi$ and to stay outside
with probability $\phi$. Since a firm entering the contest secures the contract
with certainty, the firm's probability of securing the contract is
$\mathbb{P}(\Gamma=1)=$ $1-\phi$ and not securing the contract is $\mathbb{P}(\Gamma=0)=$
$\phi$. We consider implications of the random acquisition of contract in the online appendix.

We assume that the firm's non-extortionary bribe to $B1$ can be expressed as a
fraction $\kappa\in\lbrack0,1]$ of the net payoff $V$, so that the firm transfers $\kappa V$ to $B1$ as the non-extortionary bribe and retains the payoff of $(1-\kappa)V$
upon securing the contract. Note that $B1$ does not need to observe the firm's type to elicit the bribe offer, as the bribe size is fully determined by the competition among rivaling firms in the contest (as in \cite{Lui1985} and \cite{Compte2005}) or is set as a fixed fraction of $V$.

Consistently with the empirical evidence of positive participation costs in auctions in various industries (\cite{Hortacsu2021}), we assume that participation in the contest is costly.  The firm needs to pay a fixed participation (entry) cost $gn$, where $n$ is the number of days spent in the
contest and $g>0$ is the firm's daily participation cost. Following \cite{Lui1985} and \cite{LambertMogilianskiMajumdarRadner2007}, we assume that $gn$ is sunk and cannot be recovered once the firm has entered the contest. Given the
participation cost $gn$ and the non-extortionary bribe $\kappa V$ to $B1$, the
firm's total cost of securing the contract is

\begin{equation}\label{eq:NBC_Gamma1_prelim}
NBC_{i}(\Gamma=1)=gn+\kappa V
\end{equation}

where $NBC_{i}$ stands for the non-extortionary bribe cost of the $i$-type
firm. 

We assume that the $N$-type firm cannot participate in the contest, because it
cannot afford to pay the participation cost $gn$ upfront. Evidence for this assumption can be found in the literature on both auctions (\cite{Hortacsu2021})
and corruption (\cite{Gauthier2021}), which show that the entry in auctions in general
and bribe contests in particular is endogenous due to positive entry costs,
and firms that enter such contests are more productive and profitable than those that do not enter. 

Given that the firm can have only two types and only the $P$-type firm can enter the contest, the
participation in the contest and the bribe transfer to $B1$ fully reveal the firm's type $i$ to $B1$. Since
only the $P$-type firm can afford to participate in the contest, $B1$
correctly concludes that any participating firm must be the $P$-type firm.
Further, upon allocating the contract to the firm for a bribe, $B1$ concludes that the
participating firm's profit must be larger by $(1-\kappa)V$. Hence, $B1$ knows
that the profit of the participating firm upon securing the contract is
\begin{equation}\label{eq:Pi_P_Gamma1}
\Pi_{P}(\Gamma=1)=\Pi_{N}+\Delta\Pi(\Gamma=0)+(1-\kappa)V
\end{equation}
while the profit of a firm that does not enter the contest is either $\Pi_N$ or 
\begin{equation}\label{eq:Pi_P_Gamma0}
\Pi_{P}(\Gamma=0)=\Pi_{N}+\Delta\Pi(\Gamma=0)
\end{equation} 

We note that if $B1$ transmits all the available information about the firm to $B2$, then
$B2$ correctly observes the type of the $P$-type firm with the contract but
not the type of the $P$-type firm without the contract. This leaves some
uncertainty about the $P$-type firm's type even in the presence of full
communication across periods.

Lastly, to simplify exposition, we express the non-extortionary bribe $\kappa
V$ as a fraction of the $P$-type firm's profit, $\kappa V=\lambda\Pi
_{P}(\Gamma=1)$, where $\lambda>0$ measures the size of the non-extortionary
bribe as a fraction of the $P$-type firm's profit. Then, the non-extortionary bribe cost in equation (\ref{eq:NBC_Gamma1_prelim}) is 
\begin{equation}\label{eq:NBC_Gamma1}
NBC_{i}(\Gamma=1)=gn+\lambda\Pi_{P}(\Gamma=1)
\end{equation}

\paragraph*{Extortionary bribery in period 2:}

The firm applies for a certificate from bureaucrat $B2$ in period 2. We can
think of the certificate as an approval of a tax report that all firms with
tax registration are required to possess independently of the acquisition of the contract in period 1. The number of certificates is
unlimited, and the firm meets all certificate requirements. The official
certificate fee is $F_{L}$, and it takes $t_{min}\geq0$ days for $B2$ to issue
the certificate according to the law. 

$B2$ is free to
charge an arbitrary certificate fee $F_{i,\Gamma}\in\lbrack F_{L},\infty)$ and
choose any amount of time $t_{i,\Gamma}\in\lbrack t_{min},\infty)$ to issue
the certificate.  Extortion occurs when $B2$'s choice of
$F_{i,\Gamma}$ exceeds $F_{L}$, resulting in the extortionary bribe
\begin{equation}\label{eq:EBi}
EB_{i,\Gamma}=F_{i,\Gamma}-F_{L}
\end{equation}
We assume that the $N$-type firm does not pay any extortionary bribe, so that $EB_{N}=F_{N}-F_{L}=0$. This assumption is consistent with empirical evidence,
because the median aggregate bribe of firms with non-positive profits in the data is zero. We relax this assumption in Section \ref{subsec:derivation_emp_specif} and test it empirically.

Bureaucratic delay (red tape) occurs when $B2$'s choice of
processing time $t_{i,\Gamma}$ exceeds $t_{min}$, leading to the bureaucratic
delay 
\begin{equation}\label{eq:mi}
m_{i,\Gamma}=t_{i}-t_{min}
\end{equation}
Consistently with the definition of extortion in \cite{Ayres1997}, $B2$ initiates a dispute with the firm, demands an extortionary bribe and uses bureaucratic delay at will during the dispute. 

$B2$'s utility from issuing the certificate to the $i$-type firm is%
\begin{equation}\label{eq:B2_util}
W+(F_{i,\Gamma}-F_{L})-c(t_{i,\Gamma})
\end{equation}
where $W$ is $B2$'s official salary, $F_{i,\Gamma}-F_{L}$ is the extortionary
bribe paid by the $i$-type firm, and $c(t_{i,\Gamma})$ is $B2$'s cost of
spending $t_{i,\Gamma}$ amount of time processing the certificate. We  assume that $c(t_{i,\Gamma})$ is of the form 
$c(t_{i,\Gamma}):=\frac{c}{t_{i,\Gamma}}$ for some $c>0$, so that $B2$'s
cost of processing time $c(t_{i,\Gamma})$ is positive and decreasing in the
processing time $t_{i,\Gamma}$. 

The firm's net payoff from getting the certificate is linear in $t_{i,\Gamma}$ and $F_{i,\Gamma}$ and is given by
\begin{equation}\label{eq:firm_util}
U_{i,\Gamma}  :=\Pi_{i}(\Gamma)-F_{i,\Gamma}-s_{i}t_{i,\Gamma}
\end{equation}
where $s_{i}$ is the $i$-type firm's daily cost of time. The term $\Pi_i(\Gamma)$ can take the value of $\Pi_N$, the value of $\Pi_P(\Gamma=0)$ given by (\ref{eq:Pi_P_Gamma0}), and the value of $\Pi_P(\Gamma=1)$ given by  (\ref{eq:Pi_P_Gamma1}). 

Expressing $F_{i,\Gamma}$ and $t_{i,\Gamma}$ in terms of the extortionary bribe and delay from equations (\ref{eq:EBi}) and (\ref{eq:mi}), respectively, the firm's net payoff  from getting the certificate is 
\begin{equation}\label{eq:net_payoff_period2}
U_{i,\Gamma} =\Pi_{i}(\Gamma)-F_{L}-s_{i}t_{min}-s_{i}m_{i,\Gamma}-EB_{i,\Gamma}
\end{equation}
Note that because $gn$ is sunk, it does not enter the firm's net payoff $U_{i,\Gamma}$. 

We assume that a firm with larger profit values
the certificate more ($U_{P,\Gamma=1}>U_{P,\Gamma=0}>U_{N,\Gamma}$), the daily
cost of time is positive and smaller for the $N$-type firm $(s_{P}>s_{N}>0)$,
and the daily cost of time of the $N$-type firm is not too large ($s_{N}%
<\frac{(\Pi_{N}-M-F_{N})^{2}}{c}$).

If the firm does not get the certificate, it exits the market by selling off
its assets. Hence, the firm's outside option in period $2$ is the market value
of capital $M$. 

The $i$-type firm's payoff after period $2$ is $U_{i,\Gamma} - (1-\phi)gn$, which is the difference between the net payoff after obtaining the certificate in period 2 ($U_{i,\Gamma}$ given by equation (\ref{eq:net_payoff_period2})) and the cost of participation in the contest in period 1 ($gn$) weighted by the participation probability $(1-\phi)$. Hence, the $i$-type firm's payoff after two periods is
$\Pi_{i}(\Gamma)-F_{L}-s_{i}t_{min}-(1-\phi)gn-s_{i}m_{i,\Gamma}-EB_{i,\Gamma
}$.

\subsection{Three informational scenarios} 
We consider three informational scenarios. In the no secrecy scenario (NS), $B2$ knows the firm's type and gets a report about the firm's entry into the contest from $B1$. In the secrecy with communication scenario (SC), $B2$ does not know the firm's type but $B1$ reports the firm's action to $B2$. In the secrecy without communication scenario (SwC), $B2$ neither knows the firm's type nor gets a report from $B1$. The interaction between $B1$ and $B2$ does not involve any strategic game and is limited only to the transmission of information about the firm's action in period 1. B2 faces the following events:
\begin{itemize}
\item[$I1$:] $N$-type firm requests the certificate;
\item[$I2$:] $P$-type firm without the contract from period 1 requests the certificate;
\item[$I3$:] $P$-type firm with the contract from period 1 requests the certificate.
\end{itemize}
The NS scenario is modeled as a perfect information game, where $B2$ can distinguish among these three events. The SC scenario is modeled as an imperfect information game, where $B2$ can distinguish between event $I3$ and the remaining two events, but cannot distinguish between events $I1$ and $I2$.
The SwC scenario is modeled as an imperfect information game, where $B2$ cannot distinguish among the three events. 

\subsubsection{No secrecy (NS)}\label{subsec:NS}

In the NS scenario, $B2$ observes the firm's type and action in period
$1$ and acts as a perfectly price-discriminating monopolist.  In this scenario, $B2$ distinguishes between three singleton information sets $\{ I1\}, \{ I2\}$, and $\{I3\}$, and makes three (processing time, certificate fee) take-it-or-leave-it offers: $(t_{N},F_{N})$ for the $N$-type firm in $\{I1\}$, $(t_{P,\Gamma=0},F_{P,\Gamma=0})$ for the $P$-type firm without the contract
in $\{I2\}$, and $(t_{P,\Gamma=1},F_{P,\Gamma=1})$ for the $P$-type firm with
the contract in $\{I3\}$. 

Note that since the $N$-type firm cannot pay any extortionary bribe, in the information set $\{I1\}$ we have $F_N = F_L$. The rest of the values of interest $t_N, t_{P,\Gamma=0}$,  $F_{P,\Gamma=0}, t_{P,\Gamma=1}, F_{P,\Gamma=1}$ are the optimal solutions to the problem \ref{eq:NS} below. 

In the problem \ref{eq:NS}, $B2$ maximizes his utility (\ref{eq:B2_util}) subject to the $i$th-type firm's participation constraint $IR_{i,\Gamma}$: $U_{i,\Gamma} \geq M$. For the objective function, recall that $\theta$ denotes the probability of the $N$-type firm, and $\phi$
denotes the probability of not entering the bribe contest (and not getting the certificate) in
period $1$. These probabilities weigh the terms in the $B2$'s objective (\ref{eq:B2_util}), resulting in the optimization problem \ref{eq:NS}:

\begin{align} 
\max_{(t_{N},t_{P,.},F_{P,.})} \quad & W-\theta\frac{c}{t_{N}}+(1-\theta)\phi\left(F_{P,\Gamma=0}-F_{L}-\frac
{c}{t_{P,\Gamma=0}}\right) \notag \\ 
\quad & +(1-\theta)(1-\phi)\left(F_{P,\Gamma=1}-F_{L}-\frac{c}{t_{P,\Gamma=1}}\right) \notag \\
\text{subject to } \quad & IR_{N}:\Pi_{N}-s_{N}t_{N}-F_{L}\geq M \notag \\
 \quad & IR_{P,\Gamma=0}:\Pi_{P}(\Gamma=0)-s_{P}t_{P,\Gamma=0}-F_{P,\Gamma=0}\geq M \notag \\ 
 \quad & IR_{P,\Gamma=1}:\Pi_{P}(\Gamma=1)-s_{P}t_{P,\Gamma=1}-F_{P,\Gamma=1}\geq M \label{eq:NS} \tag{NS}
\end{align}

where for notational convenience $t_{P,.}$ denotes optimization variables $t_{P,\Gamma=0}$ and $t_{P,\Gamma=1}$, while $F_{P,.}$ denotes variables $F_{P,\Gamma=0}$ and $F_{P,\Gamma=1}$.

\paragraph*{Solution to \ref{eq:NS}:} $B2$ can strictly raise his payoff by increasing $F_{P,\Gamma=0}$,
$F_{P,\Gamma=1}$, and $t_{N}$ until each participation constraint is binding.
Optimization with binding $IR_{P,\Gamma=0}$ and $IR_{P,\Gamma=1}$ yields the
following, with the superscript $NS$ indicating the $NS$ scenario:

\begin{gather}\label{eq:sol_NS_tN}
t_{N}^{NS}=\frac{\Pi_{N}-M-F_{L}}{s_{N}} \\\label{eq:sol_NS_FN}
F_{N}^{NS}=F_{L} \\\label{eq:sol_NS_tP} 
t_{P}^{NS}=t_{P,\Gamma=0}^{NS}=t_{P,\Gamma=1}^{NS}=\sqrt{\frac{c}{s_{P}}}\\\label{eq:sol_NS_FP0}
F_{P,\Gamma=0}^{NS}=\Pi_{P}(\Gamma=0)-M-s_{P}t_{P}^{NS}\\ \label{eq:sol_NS_FP1}
F_{P,\Gamma=1}^{NS}=\Pi_{P}(\Gamma=1)-M-s_{P}t_{P}^{NS} 
\end{gather}

\paragraph*{Equilibrium bribes:} Since the $N$-type firm cannot pay any bribe and $B2$'s increase in utility
from serving the $N$-type firm only comes from taking longer to issue the
certificate, $B2$ sets the certificate processing time for the $N$-type firm
at the highest possible level, while satisfying the $N$-type firm's
participation constraint. This yields the $N$-type firm's extortionary bribe
of $EB_{N}^{NS}=F_{N}^{NS}-F_{L}=0$ and bureaucratic delay of 
$m_{N}^{NS}=t_{N}^{NS}-t_{min}$, where $t_N^{NS}$ is given by equation (\ref{eq:sol_NS_tN}). 
As the $N$-type firm does not enter the bribe contest in period $1$ and has zero
non-extortionary bribe cost ($NBC_{N}=0$), the $N$-type firm's aggregate bribe
$Bribe_{N}^{NS}$ is identical to the extortionary bribe $EB_{N}^{NS}$, i.e. $Bribe_{N}^{NS}=EB_{N}^{NS}=0$.

For the $P$-type firm with or without the contract, the take-it-or-leave-it certificate fees are given by 
$EB_{P,\Gamma=0}^{NS}=F_{P,\Gamma=0}^{NS}-F_{L}$ and $EB_{P,\Gamma=1}^{NS}=F_{P,\Gamma=1}^{NS}-F_{L}$
respectively. Note that the extortionary bribe of the $P$-type firm with the contract is larger, since $F_{P,\Gamma=1}^{NS} > F_{P,\Gamma=0}^{NS}$. Solving
backwards, the $P$-type firm's dominant strategy is not to enter the bribe
contest in period $1$ ($\phi=1$). Hence, the $P$-type firm's equilibrium extortionary
bribe is $EB_{P}^{NS} = F_{P,\Gamma=0}^{NS}-F_{L}$,
where $F_{P,\Gamma=0}^{NS}$ is given by equation (\ref{eq:sol_NS_FP0}). Letting $m_P^{NS} = t_P^{NS} - t_{min}$ to denote the firm's processing delay, the $P$-type firm's aggregate bribe in equilibrium is
\begin{equation}\label{eq:eq_bribe_P_NS}
Bribe_{P}^{NS}=EB_{P}^{NS}=\Pi_{N}+\Delta\Pi(\Gamma=0)-M-F_{L}-s_{P}
t_{min}-s_{P}m_{P}^{NS}
\end{equation}
Proposition \ref{prop:eq_bribe_NS} reports equilibrium aggregate bribes in the NS scenario:

\begin{proposition} [Bribery in the no secrecy (NS) scenario] \label{prop:eq_bribe_NS}

If $B2$ is perfectly informed about the firm's type and action in period $1$, then 
\begin{itemize}
\item[(a)] The $N$-type firm pays aggregate bribe $Bribe_{N}^{NS}=0,$
\item[(b)] The $P$-type firm never enters the bribe contest in period $1$ ($\phi=1$) and pays aggregate bribe $Bribe_{P}^{NS}$ given by equation (\ref{eq:eq_bribe_P_NS}).
\end{itemize}
\end{proposition}

\subsubsection{Secrecy with communication (SC)}\label{subsec:SC}

In the SC scenario, $B2$ does not observe the firm's type, but $B1$ informs $B2$ about the
firm's entry into the bribe contest in period $1$. In this scenario, $B2$ has two information sets: $\{I1,I2\}$ and $\{I3\}$. 

In the information set $\{I3\}$, $B2$ perfectly observes the firm's type and profit (the $P$-type firm with
the contract), makes take-it-or-leave-it offer $(t_{P,\Gamma=1},F_{P,\Gamma=1})$ to the firm, and fully extracts
the firm's surplus. The certificate processing time and fee of the
$P$-type firm with the contract coincide with those in the NS scenario: $t_{P,\Gamma=1}^{SC}=t_{P,\Gamma=1}^{NS}$ and
$F_{P,\Gamma=1}^{SC}=F_{P,\Gamma=1}^{NS}$
(the superscript $SC$ indicates the $SC$ scenario).

In the information set $\{I1,I2\}$, $B2$ faces the request for certificate from the firm of unknown type 
that did not enter the bribe contest in period $1$. Nevertheless, $B2$ knows that with probability $\frac
{\theta}{\theta+(1-\theta)\phi}>0$ the request comes from the $N$-type firm
and with probability $\frac{(1-\theta)\phi}{\theta+(1-\theta)\phi}$ the request comes from the
$P$-type firm without the contract. Hence, in the information set\thinspace$\{I1,I2\}$,
$B2$ offers a menu of $(t_{N},F_{N})$ and $(t_{P,\Gamma=0},F_{P,\Gamma=0})$, where $F_{N}=F_{L}$. The rest of the values $t_N$, $t_{P,\Gamma=0}$, and $F_{P,\Gamma=0}$ are the optimal solutions to the screening problem \ref{eq:SC} below. 

The objective in the screening problem \ref{eq:SC} is to maximize $B2$'s utility (\ref{eq:B2_util}), with terms representing contributions from the $N$-type firm and the $P$-type firm without the contract weighted by their respective probabilities. The participation constraints $IR_N$ and $IR_{P,\Gamma=0}$ and the incentive compatibility constraints $IC_{N}$ and $IC_{P,\Gamma=0}$ for the $N$-type firm and the $P$-type firm without the contract, respectively, use expression for $U_{i,\Gamma}$ given by equation (\ref{eq:firm_util})). 
\small{
\begin{align} 
\max_{(t_{N},t_{P,\Gamma=0},F_{P,\Gamma=0})} \quad & W-\frac{\theta}
{\theta+(1-\theta)\phi}\frac{c}{t_{N}}+\frac{(1-\theta)\phi}{\theta
+(1-\theta)\phi}\left(F_{P,\Gamma=0}-F_{L}-\frac{c}{t_{P,\Gamma=0}}\right) \notag \\
\text{ subject
to } \quad & IR_{N}:\Pi_{N}-s_{N}t_{N}-F_{L}\geq M \notag \\
\quad & IR_{P,\Gamma=0}:\Pi_{P}(\Gamma=0)-s_{P}t_{P,\Gamma=0}-F_{P,\Gamma=0}\geq M \notag\\
\quad & IC_{N}:\Pi_{N}-s_{N}t_{N}-F_{L}\geq\Pi_{N}-s_{N}t_{P,\Gamma=0}-F_{P,\Gamma
=0} \notag\\
\quad & IC_{P,\Gamma=0}:\Pi_{P}(\Gamma=0)-s_{P}t_{P,\Gamma=0}-F_{P,\Gamma=0} \geq \Pi_{P}(\Gamma=0)-s_{P}t_{N}-F_{L} \label{eq:SC} \tag{SC} 
\end{align}
} 
\normalsize The solution to the screening problem \ref{eq:SC} is given in Lemma \ref{lem:SC}. The proof of the lemma uses  routine arguments and is relegated to the online appendix. 

\begin{lemma}[Solution to the screening problem \ref{eq:SC}]\label{lem:SC}
In the information set $\{I1,I2\}$, $B2$ offers processing times and  certificate fees: $t_{N}^{SC}=\frac{\Pi_{N}-M-F_{L}}{s_{N}}$ and $F_{N}^{SC}=F_{L}$ to the $N$-type firm (same as in the NS case) and 
$t_{P,\Gamma=0}^{SC}=\sqrt
{\frac{c}{s_{P}}}$ and $F_{P,\Gamma=0}^{SC}=s_{P}(t_{N}^{SC}
-t_{P,\Gamma=0}^{SC})+F_{L}$ to the $P$-type firm.
\end{lemma}

\paragraph*{Equilibrium bribes:} Since the processing time for the $N$-type firm is the same and no
extortionary bribe is requested, the aggregate bribe of the $N$-type firm in this scenario is the same as in the NS scenario: $Bribe_{N}^{SC}=Bribe_{N}^{NS}=0$.

Because of the $P$-type firm's information rent in the information set $\{I1,I2\}$, the payoff of the $P$-type firm without the
contract strictly exceeds the payoff of the $P$-type firm with the contract,
whose type is fully revealed by $B2$'s message to $B1$. By
backward induction, we conclude that the $P$-type firm never obtains the
contract in period $1$ ($\phi=1$), and the $P$-type firm's equilibrium
extortionary bribe is $EB_{P}^{SC}=F_{P,\Gamma=0}^{SC}-F_{L}$,
which by Lemma \ref{lem:SC} is $F_{P,\Gamma=0}^{SC}-F_{L} = s_{P}\left(t_{N}^{SC}-t_{P,\Gamma
=0}^{SC}\right)$. Since by Lemma \ref{lem:SC}, $ t_{N}^{SC}=\frac{\Pi_{N}-M-F_{L}}{s_{N}}$ and  $m_P^{SC}=t_{P,\Gamma=0}^{SC} - t_{min}$, we obtain 

\begin{equation}\label{eq:eq_bribe_P_SC}
EB_P^{SC}=\frac{s_{P}}{s_{N}}(\Pi_{N}-M-
F_{L})-s_{P}t_{min}-s_{P}m_{P}^{SC}
\end{equation}
Since $\phi=1$, the $P$-type never enters the bribe
contest in period $1$, and the $P$-type firm's aggregate bribe is equal to the
extortionary bribe $Bribe_{P}^{SC}=EB_{P}^{SC}>0$. Note that $Bribe_{P}^{SC}$ in this scenario does not depend on the $P$-type firm's actual
profit $\Pi_{P}$; it is completely determined by waiting costs and
bureaucratic delays. 

Proposition \ref{prop:eq_bribe_SC} summarizes these findings and reports equilibrium aggregate bribes in the secrecy with communication (SC) scenario:
\begin{proposition}[Bribery in the secrecy with communication (SC) scenario]\label{prop:eq_bribe_SC}\hfill
If $B2$ is uninformed about the firm's type but informed about the firm's
action in period $1$, then
\begin{itemize}
\item[(a)] The $N$-type firm pays aggregate
bribe $Bribe_{N}^{SC}=0,$
\item[(b)] The $P$-type firm never
enters the bribe contest in period $1$ ($\phi=1$) and pays aggregate
bribe $Bribe_P^{SC}$ given by equation (\ref{eq:eq_bribe_P_SC}).
\end{itemize}
\end{proposition}

\subsubsection{Secrecy without communication (SwC)}\label{subsec:SwC}

In the SwC scenario, $B2$ does not know the firm's type, and $B1$ does not inform $B2$ about
the firm's participation in the bribe contest in period $1$. $B2$ has the single information set $\{I1,I2,I3\}$, in which he cannot distinguish among requests for the certificate from firms of different types. In this information set, $B2$ knows that the request for certificate comes from the
$N$-type firm with probability $\theta$, from the $P$-type firm without the
contract with probability $(1-\theta)\phi$, and from the $P$-type firm with the
contract with probability $(1-\theta)(1-\phi)$. 

In the single information set, $B2$ offers a menu of three
certificate processing times and fees:
$(t_{N},F_{N})$ for the $N$-type firm, $(t_{P,\Gamma=0},F_{P,\Gamma=0})$ for
the $P$-type firm without the contract, and $(t_{P,\Gamma=1},F_{P,\Gamma=1})$
for the $P$-type firm with the contract. We have $F_{N}=F_{L}$, as the $N$-type firm cannot pay any bribe. The rest of the values $t_N$, $t_{P,\Gamma=0}$, $F_{P,\Gamma=0}$, $t_{P,\Gamma=1}$, $F_{P,\Gamma=1}$ are the optimal solutions to the screening problem \ref{eq:SwC} below.

As before, the objective function in the screening problem is based on (\ref{eq:B2_util}), and constraints are based on (\ref{eq:firm_util}). $IR_{i,\Gamma}$ denotes the $i$-type firm's
participation constraint. $IC_{N}(P,\Gamma)$ denotes the incentive
compatibility constraint of the $N$-type firm imitating the $P$-type firm with or without the contract. $IC_{P,\Gamma=0}(N)$ and $IC_{P,\Gamma=0}(P,\Gamma=1)$ denote the incentive compatibility constraints of the $P$-type firm without the contract imitating the $N$-type firm and the $P$-type firm with the contract, respectively. $IC_{P,\Gamma=1}(N)$ and $IC_{P,\Gamma=1}(P,\Gamma=0)$ denote the incentive compatibility constraints of the $P$-type firm with the contract imitating the $N$-type firm and the $P$-type firm without the contract, respectively. The \ref{eq:SwC} problem reads
\footnotesize{ 
\begin{align} 
\max_{(t_{N},t_{P,.},F_{P,.})} \quad & W-\theta\frac{c}{t_{N}}+\phi(1-\theta)\left(F_{P,\Gamma=0}-F_{L}-\frac
{c}{t_{P,\Gamma=0}}\right)+(1-\phi)(1-\theta)\left(F_{P,\Gamma=1}-F_{L}-\frac{c}{t_{P,\Gamma=1}}\right) \notag \\
\text{
subject to} \quad & IR_{N}:\Pi_{N}-s_{N}t_{N}-F_{L}\geq M \notag \\
 \quad & IR_{P,\Gamma=0}:\Pi_{P}(\Gamma=0)-s_{P}t_{P,\Gamma=0}-F_{P,\Gamma=0}\geq M \notag \\
 \quad & IR_{P,\Gamma=1}:\Pi_{P}(\Gamma=1)-s_{P}t_{P,\Gamma=1}-F_{P,\Gamma=1}\geq M \notag \\
\quad & IC_{N}(P,\Gamma=0):\Pi_{N}-s_{N}t_{N}-F_{L}
\geq\Pi_{N}-s_{N}t_{P,\Gamma
=0}-F_{P,\Gamma=0} \notag \\
\quad & IC_{N}(P,\Gamma=1):\Pi_{N}-s_{N}t_{N}-F_{L} 
\geq\Pi_{N}-s_{N}t_{P,\Gamma
=1}-F_{P,\Gamma=1} \notag \\
\quad & IC_{P,\Gamma=0}(N):\Pi_{P}(\Gamma=0)-s_{P}t_{P,\Gamma=0}-F_{P,\Gamma=0} 
\geq \Pi_{P}(\Gamma=0)-s_{P}t_{N}-F_{L} \notag \\
\quad & IC_{P,\Gamma=0}(P,\Gamma=1):\Pi_{P}(\Gamma=0)-s_{P}t_{P,\Gamma=0}
-F_{P,\Gamma=0}
\geq\Pi_{P}(\Gamma=0)-s_{P}t_{P,\Gamma=1}-F_{P,\Gamma=1} \notag \\
\quad & IC_{P,\Gamma=1}(N):\Pi_{P}(\Gamma=1)-s_{P}t_{P,\Gamma=1}-F_{P,\Gamma=1} \geq \Pi_{P}(\Gamma=1)-s_{P}t_{N}-F_{L} \notag \\
\quad & IC_{P,\Gamma=1}(P,\Gamma=0):\Pi_{P}(\Gamma=1)-s_{P}t_{P,\Gamma=1}
-F_{P,\Gamma=1} \geq\Pi_{P}(\Gamma=1)-s_{P}t_{P,\Gamma=0}-F_{P,\Gamma=0} \label{eq:SwC}\tag{SwC} 
\end{align}
}
\normalsize where for notational convenience $t_{P,.}$ denotes variables $t_{P,\Gamma=0}$ and $t_{P,\Gamma=1}$ and $F_{P,.}$ denotes variables $F_{P,\Gamma=0}$ and $F_{P,\Gamma=1}$. The superscript $SwC$ indicates the SwC scenario.

\paragraph*{Equilibrium bribes:} The optimal processing time and fee for the $N$-type firm are the same as in the NS and SC scenarios: $t_{N}^{SwC}=t_{N}^{SC}=t_{N}^{NS}$
and $F_N^{SwC} = F_N^{SC} = F_N^{NS} = F_L$. This is because the $N$-type firm always faces the longest possible processing time that is determined by its binding participation constraint $IR_{N}$. 

Since the $P$-type firm's cost of time $s_{P}$ does
not depend on the firm's entry into the bribe contest, $B2$ cannot use certificate processing time to discriminate between the
$P$-type firm with and without the contract. Hence, the menu offers for the $P$-type firm with and without the contract in
the information set $\{I1,I2,I3\}$ are identical: $t_{P,\Gamma=0}^{SwC}=t_{P,\Gamma=0}^{SC}=t_{P,\Gamma=0}^{NS}=t_{P,\Gamma=1}^{SwC}=t_{P,\Gamma=1}^{SC}=t_{P,\Gamma=1}^{NS}$
with the value given in equation (\ref{eq:sol_NS_tP}), and 
$F_{P,\Gamma=1}^{SwC}=F_{P,\Gamma=0}^{SwC}=F_{P,\Gamma=0}^{SC}$ as given by Lemma \ref{lem:SC}.

Solving backwards, the $P$-type firm enters the bribe contest ($\phi=0$), because it obtains a larger profit while paying
the same extortionary bribe as without the contract. For this reason, $B2$
concludes that in $\{I1,I2,I3\}$ the firm can have only two profit
levels: $\Pi_{N}$ and $\Pi_{P}(\Gamma=1)$. Since the solution to the screening
problem with two firm types does not depend on the $P$-type firm's actual
profit, the certificate fee in the optimal menu
for the $P$-type firm with the contract in this scenario  must be
identical to the one offered to the $P$-type firm in the SC scenario (Lemma \ref{lem:SC}). 

The solution to the \ref{eq:SwC} problem is presented in Lemma \ref{lem:SwC}. The proof uses routine techniques and is relegated to the online appendix.

\begin{lemma}[Solution to the screening problem \ref{eq:SwC}] \label{lem:SwC}
In the information set $\{I1,I2,I3\}$, in equilibrium the
$N$-type firm faces certificate processing time $t_{N}^{SwC}=\frac
{\Pi_{N}-M-F_{L}}{s_{N}}$ and pays fee $F_{N}^{SwC}=F_{L}$. The $P$-type firm enters the bribe contest in period
$1$ ($\phi=0$),  faces certificate processing time $t_{P}^{SwC}
=\sqrt{\frac{c}{s_{P}}}$ and pays fee $F_{P}^{SwC}=s_{P}(t_{N}^{SwC}-t_{P}^{SwC})+F_{L}$.
\end{lemma}

The aggregate bribe and delay of the $N$-type firm in this scenario remain the same as in the NS and SC scenarios: $Bribe_{N}^{SwC}=Bribe_{N}^{SC}=Bribe_{N}^{NS}=0$ and  
$m_{N}^{SwC}=m_{N}^{SC}=m_{N}^{SC}=\frac{\Pi_{N}-M-F_{L}}{s_{N}}-t_{min}$. The delay and the extortionary bribe of the $P$-type firm are identical to those in the SC scenario: $m_{P}^{SwC}=m_{P}^{SC}$ and $EB_{P}^{SwC}=EB_{P}^{SC}$
as in (\ref{eq:eq_bribe_P_SC}). However, since the $P$-type firm enters the bribe contest ($\phi=0$), it has positive
non-extortionary bribe cost $NBC_{P}(\Gamma=1)=gn+\lambda\Pi_{P}(\Gamma=1),$ yielding the equilibrium aggregate bribe
\begin{align}\notag
Bribe_{P}^{SwC}  &  =EB_{P}^{SwC}+NBC_{P}(\Gamma=1)\\\label{eq:eq_bribe_P_SwC}
&  =\frac{s_{P}}{s_{N}}(\Pi_{N}-M-
F_{L})-s_{P}t_{min}-s_{P}m_{P}^{SwC}+gn+\lambda\Pi_{P}(\Gamma=1)
\end{align}
Proposition \ref{prop:eq_bribe_SwC} reports equilibrium aggregate bribes in the SwC scenario:

\begin{proposition}[Bribery in the secrecy without communication (SwC) scenario]\label{prop:eq_bribe_SwC}
If $B2$ is uninformed about the firm's type and action in period $1$, then 
\begin{itemize}
\item[(a)] The $N$-type firm pays aggregate bribe $Bribe_{N}^{SwC}=0$
\item[(b)] The $P$-type firm pays aggregate bribe given by equation (\ref{eq:eq_bribe_P_SwC}).
\end{itemize}
\end{proposition}


\section{Empirical specification}\label{sec:identification}
The goal of the empirical analysis is (i) to test for the multiplicity of bribery types and (ii) to test for interactions between different types of bribery by establishing which informational scenario is consistent with the observed behavior of firms: NS of Section \ref{subsec:NS}, SC of Section \ref{subsec:SC}, or SwC of Section \ref{subsec:SwC}. 

\subsection{Derivation of the empirical specification}\label{subsec:derivation_emp_specif} 
To construct the empirical specification, we use the following indicator variables:
\begin{itemize}
\item $1_{disp}$: an indicator of bureaucratic dispute that serves as a proxy for extortionary bribe request;
 
\item $1_N$, $1_P$: indicators for $N$-type and $P$-type firms, where $1_N+1_P=1$;
 
\item $1_{NS}$, $1_{SC}$, $1_{SwC}$: indicators for the NS, SC, and SwC scenarios, respectively, where $1_{NS} + 1_{SC} + 1_{SwC} = 1$.

\end{itemize}

If the $N$-type firm could pay a positive extortionary bribe, $B2$ would demand a certificate fee $F_{N}^{*}$ satisfying the $N$-type firm's participation constraint $IR_{N}:\Pi_{N}-s_{N}t_{N}-F_{N}\geq M$ as equality, leading to $F_{N}^{*}=\Pi_{N}-s_{N}t_{N}-M$. Then (assuming that the $N$-type firm never enters the bribe contest in period $1$) the $N$-type firm's aggregate bribe would be
$Bribe_{N}^{*}=EB_{N}^{*}=F_{N}^{*}-F_{L}=\Pi_{N}-M-F_{L}-s_{N}t_{min} -s_{N}m_{N}$.  Hence, to incorporate the possibility of
the $N$-type firm paying positive extortionary bribe in any scenario, we express the
$N$-type firm's aggregate bribe as
\begin{equation}\label{eq:bribeN}
Bribe_{N}^{*}=1_{B}(\Pi_{N}-M-F_{L}-s_{N}t_{min}-s_{N}m_{N})
\end{equation}
where $1_{B}=1$ for the $N$-type firm paying positive extortionary bribe, and $1_{B}=0$ otherwise.

Propositions \ref{prop:eq_bribe_NS}, \ref{prop:eq_bribe_SC}, and \ref{prop:eq_bribe_SwC} indicate that the $P$-type firm pays both types of bribes in the SwC scenario and only extortionary bribe in the NS and SC scenarios. Hence, the aggregate bribe in all informational scenarios for both firm types is

\[
Bribe=\left\{
\begin{array}
[c]{l}%
1_{B}(\Pi_{N}-M-F_{L}-s_{{\small N}}t_{min}-s_{{\small N}}m_{N})1_{disp}%
\text{,\ if }1_{N}=1 \text{ from } (\ref{eq:bribeN}) \\ 
\left(\Pi_{N}+\Delta\Pi(\Gamma=0)-M-F_{L}-s_{P}t_{min}-s_{P}m_{P}\right)1_{disp} %
\text{,\ if } 1_{P}=1\text{ \& }1_{NS}=1 \text{ from (\ref{eq:eq_bribe_P_NS})}\\
\left(\frac{s_{P}}{s_{N}}(\Pi_{N}-M-F_{L})-s_{P}t_{min}-s_{P}m_{P}\right)1_{disp}%
\text{,\ if
}1_{P}=1\text{ \& }1_{SC}=1 \text{ from (\ref{eq:eq_bribe_P_SC})}\\

\left(\frac{s_{P}}{s_{N}}(\Pi_{N}-M-F_{L})-s_{P}t_{min}-s_{P}m_{P}\right)1_{disp}%
+gn+\lambda\Pi_{P}(\Gamma=1)\text{, }\\
\text{
\ \ \ \ \ \ \ \ \ \ \ \ \ \ \ \ \ \ \ \ \ \ \ \ \ \ \ \ \ \ \ \ \ \ \ \ \ \ \ \ \ \ \ \ \ \ \ \ \ \ \ \ \ \ \ \ \ \ \ \ \ \ \ \ if
}1_{P}=1\text{ \& }1_{SwC}=1 \text{ from (\ref{eq:eq_bribe_P_SwC})}\\  
\end{array}
\right.
\]

After regrouping variables, we obtain the following expression:

\small{
\begin{align*}
\text{{}}Bribe  &  =1_{NS}\Delta\Pi(\Gamma=0)1_{P}1_{disp}+1_{SwC}\lambda
\Delta\Pi(\Gamma=1)1_{P}\\
&  +\underbrace{\left[\left(1_{B}1_{N}+1_{P}\frac{s_{N}1_{NS}+s_{P}(1-1_{NS})}{s_{N}}\right)1_{disp} +1_{P}1_{SwC}\lambda\right]}_{\beta_{\Pi_N}}\Pi_{N}\\
& \underbrace{-\left(1_{B}1_{N}+1_{P}\frac{s_{N}1_{NS}+s_{P}(1-1_{NS})}{s_{N}}\right)}_{\beta_M = \beta_{F_L}}M1_{disp} \underbrace{-\left(1_{B}1_{N}+1_{P}\frac{s_{N}1_{NS}+s_{P}(1-1_{NS})}{s_{N}}\right)}_{\beta_M = \beta_{F_L}}F_{L}1_{disp}\\
& \underbrace{ -\left(s_{{\small N}}1_{B}1_{N}+s_{{\small P}}1_{P}\right)}_{\beta_{t_{min}}}t_{min}1_{disp}%
-s_{{\small N}}1_{B}\underbracket{m_{N}1_{N}1_{disp}}_{m_N}-s_{{\small P}}\underbracket{m_{P}1_{P}1_{disp}}_{m_P}
+1_{SwC}g\underbracket{n1_{P}}_{n}
\end{align*}
}
\normalsize
where $m_N$ is the delay of the $N$-type firm during dispute, $m_P$ is the delay of the $P$-type firm during dispute, $n$ is the $P$-type firm's participation time in the bribe contest. Thus, we obtain the following specification \ref{eq:EM1}, with coefficients to be estimated included in parentheses:

\small{
\begin{align}\label{eq:EM1} \notag
Bribe  &  = (1_{NS})\Delta\Pi
(\Gamma=0)1_{P}1_{disp}+(1_{SwC}\lambda)\Delta\Pi(\Gamma=1)1_{P} + (\beta_{{\small \Pi}_{N}})\Pi_{N}+(\beta_{{\small M}})M1_{disp}\\ \nonumber
 & +(\beta_{{\small F}_{L}})F_{L}1_{disp}+(\beta_{{\small t}_{min}})%
t_{min}1_{disp}-(s_{{\small N}}1_{B})m_{N}-(s_{{\small P}})m_{P}+(1_{SwC}%
g)n \tag{EM1}
\end{align}
}
\normalsize

\subsection{Identification}\label{subsec:identification}

\begin{table}
\center{
\caption{Identification of informational scenarios NS, SC, and SwC.}
\center
\small{
\begin{NiceTabular}[c]{c r r r} 
 \hline
 \hline
Variables in & Coefficients in & Coefficients in & Coefficients in \\ 
  specification \ref{eq:EM1} & NS scenario & SC scenario & SwC scenario \\
 \hline
 $\Delta\Pi(\Gamma=0)1_P 1_{disp}$ & $1_{NS}>0$ & $1_{NS}=0$ & \Block[draw=red,rounded-corners]{3-1}{} $1_{NS}=0$ \\ 
 $\Delta\Pi(\Gamma=1)1_P$ & $1_{SwC}\lambda=0$ &  $1_{SwC}\lambda=0$ & $1_{SwC}\lambda>0$ \\
 $n$ & $1_{SwC}g=0$ & $1_{SwC}g=0$ & $1_{SwC}g>0$ \\
 $\Pi_N$ & $\beta_{\Pi_N} \geq 0$ & $\beta_{\Pi_N} \geq 0$ & $\beta_{\Pi_N} \geq 0$ \\
 $M1_{disp}$ & $\beta_M \leq 0$ & $\beta_M \leq 0$ & $\beta_M \leq 0$ \\
 $F_L 1_{disp}$ & $\beta_{F_L} \leq 0$ & $\beta_{F_L} \leq 0$ & $\beta_{F_L} \leq 0$ \\
 $t_{min} 1_{disp}$ & $\beta_{t_{min}} \leq 0$ & $\beta_{t_{min}} \leq 0$ & $\beta_{t_{min}} \leq 0$ \\
 $m_N$ & $s_N 1_B \geq 0$ & $s_N 1_B \geq 0$ & $s_N 1_B \geq 0$ \\
 $m_P$ & $s_P>0$ & $s_P>0$ & $s_P>0$ \\
 \hline
 \hline
\end{NiceTabular} \label{table:ident_inf_scen}
}
 \caption*{\footnotesize{Note: An informational scenario can be identified by coefficients in the first three rows (variables $\Delta(\Gamma=0)1_P 1_{disp}$, $\Delta \Pi(\Gamma=1)1_P$, and $n$). The red box indicates the scenario consistent with our data when we estimate the empirical specification  \ref{eq:EM11} (see Table \ref{table:coefs} for the estimated coefficients).}}
 }
\end{table}

\normalsize

We use specification \ref{eq:EM1} to simultaneously identify both bribery types and three informational scenarios. We can identify all three informational scenarios only in the presence of both types of bribery. We can identify the non-extortionary bribery only in the SwC scenario. Table \ref{table:ident_inf_scen} reports restrictions on coefficients in \ref{eq:EM1} consistent with each informational scenario.

\paragraph*{Identification of non-extortionary bribery:}  The coefficients of $n$ and
$\Delta\Pi(\Gamma=1)1_P$ are the sources of identification of the non-extortionary
bribery. The coefficient of $n$ identifies $1_{SwC}g$, the $P$-type firm's daily participation cost in the bribe contest in the SwC scenario. In the presence of non-extortionary bribery with costly participation, $g$ is strictly positive.
The zero $g$ indicates negligent participation cost, and the negative $g$
is inconsistent with non-extortionary bribery. The coefficient of
$\Delta\Pi(\Gamma=1)1_P$ identifies $1_{SwC}\lambda$, the size of the non-extortionary
bribe (as a share of positive profit increment) paid to $B1$ by the firm in the bribe
contest in the SwC scenario. In the presence of non-extortionary bribery, $\lambda$ is strictly
positive. The zero $\lambda$ indicates the absence of non-extortionary bribery.

\paragraph*{Identification of extortionary bribery:} The coefficients of $m_{P}$ and $m_{N}$ are the sources of identification
of the extortionary bribery. The coefficient of $m_{P}$ identifies $-s_{{\small P}%
}$, the negative of the $P$-type firm's daily cost of time. The
coefficient of $m_{N}$ identifies $-s_N 1_B$, the negative of the daily cost of time of the $N$-type firm paying positive extortionary bribe. When $B2$ uses delay to extort bribes, these
coefficients are negative, and the coefficient of $m_{P}$ is larger in 
absolute size than the coefficient of $m_{N}$. When coefficients of $m_{P}$
and $m_{N}$ are similar in size, $B2$ cannot use bureaucratic delay for
screening and sets a
single bribe-delay combination for both firm types. Positive coefficients of
$m_{P}$ and $m_{N}$ are inconsistent with extortionary bribery. 

\paragraph*{Identification of the NS scenario:} The coefficient of $\Delta\Pi(\Gamma=0)1_{P}1_{disp}$ is the source of
identification of the NS scenario.
When $B2$ is perfectly informed about the firm's type and action, the
$P$-type firm's extortionary bribe is an
increasing function of the actual profit $\Pi_{P}$, and therefore, of both $\Pi_{N}$ and $\Delta\Pi(\Gamma=0)$. Hence, in this scenario, the coefficient of
$\Delta\Pi(\Gamma=0)1_P 1_{disp}$ is strictly positive. Further, in this scenario
the $P$-type firm never enters the bribe contest, and
coefficients of parameters relevant for the non-extortionary bribery, $\Delta
\Pi(\Gamma=1)1_{P}$ and $n$, are zero.

\paragraph*{Identification of the SC scenario:} The zero coefficients of $\Delta\Pi
(\Gamma=0)1_{P}1_{disp}$, $\Delta\Pi(\Gamma=1)1_{P}$, and $n$ identify the SC scenario. When $B2$ is
uninformed about the firm's type but receives a message about the
firm's participation in the bribe contest, the $P$-type firm
does not enter the bribe contest in equilibrium, and coefficients of both $\Delta\Pi
(\Gamma=1)1_{P}$ and $n$ are zero. Consequently, $B2$ relies only on the screening mechanism to
extort a bribe from the $P$-type firm, which means that the extortionary bribe (during dispute) depends only on $\Pi_{N}$ and not on $\Delta\Pi$. Therefore, the coefficient of $\Delta\Pi(\Gamma=0)1_{P}1_{disp}$ is zero.

\paragraph*{Identification of the SwC scenario:} Coefficients of $\Delta\Pi(\Gamma=1)1_{P}$ and $n$ are the sources of
identification of the SwC scenario. When $B2$ does not know the firm's type and action, the $P$-type firm enters the bribe
contest in period $1$, covers sunk participation cost $gn$, and pays the
non-extortionary bribe as a fraction of both $\Pi_{N}$ and $\Delta\Pi
(\Gamma=1)$. Hence, the coefficients of $\Delta\Pi
(\Gamma=1)1_{P}$ and $n$ are strictly positive. Further, since in this
scenario $B2$ relies only on the screening mechanism to extort a bribe, the
extortionary bribe depends only on $\Pi_{N}$ and not on the
positive profit increment $\Delta\Pi$. Therefore, the coefficient of
$\Delta\Pi(\Gamma=0)1_{P}1_{disp}$ is zero.

\section{Estimation}\label{sec:estimation}

\subsection{Data}\label{subsec:data}

We use data from the nationally representative survey of 429 firms operating
in Tajikistan in 2012-2014. The data were collected by an agency that
collected similar data for the World Bank, and the data collection procedure
was identical to that of WBES. There are two
differences between the data used in this paper and the data from WBES: (1)
the data in this paper contain information about additional variables (e.g.
bureaucratic disputes, bureaucratic delays, unreported profits) not found in WBES, and (2) the data in this paper are panel data rather than cross-sectional. All variables involving monetary
values were collected in local currency, adjusted for inflation, and converted
to US dollars in 2014\footnote{The data were adjusted for inflation rates of
6.076\% in 2014 and 5.035\% in 2013 (source: World Bank) and converted to US
dollars at the rate of 5 TJ somonis per 1 US dollar (exchange rate in December
2014).}. Summary statistics are reported in Table \ref{table:data_summary}. The key points of the construction of variables in equation \ref{eq:EM1} are discussed below. Detailed information on the data collection procedure and the construction of all variables is in the online appendix. 

We use profit that each firm reports in its tax report as a measure for $\Pi_{N}$, and the difference between reported profit and the profit adjusted for under-reported sales and over-stated costs as a measure for the positive
profit increment $\Delta\Pi(\Gamma)$. Firms with positive
adjusted profits are the $P$-type firms ($1_{P}=1$) and firms with non-positive adjusted profits are the $N$-type firms ($1_{N}=1_{P}-1$). 

We use two measures of aggregate bribes. The indirect measure ($Bribe$) is the difference between actual paperwork expenditures and expenditures only through official channels (measure of $F_{L}$), and it represents paperwork expenditures in excess of those required by law. 
The direct measure ($Rep\_bribe$) is the reported annual bribe as a share of annual sales (bribe measure collected by WBES). The mean difference between these two measures is US \$ -0.4, and the 95\% confidence interval is $[-13.9,13.2]$.

We use an indicator of the positive number of disputes initiated by bureaucrats ($1_{disp}$) as an indicator of extortionary bribery. To construct measures for bureaucratic delays ($m_{N},m_{P},n$), we use unspecified bureaucratic delay ($Delay$), which is the difference between the actual number of days that firms spend with bureaucrats and the aggregate paperwork processing time specified by law (measure of $t_{min}$). To measure $m_{P}$, we interact $Delay$ with the indicator of the $P$-type firm ($1_{P}$) and the indicator of bureaucratic disputes ($1_{disp}$), so that $m_{P}=Delay1_{P}1_{disp}$. To measure $m_{N}$, we interact $Delay$ with the indicator of the $N$-type firm ($1_N=1-1_{P}$) and the indicator of bureaucratic disputes, $m_{N}=Delay(1-1_{P})1_{disp}$. To measure $n$, we interact $Delay$ with the indicator of the $P$-type firm, $n=$ $Delay1_{P}$.

\begin{table}
\caption{Summary statistics (all years in 2014 US \$)}.
\small{
\begin{tabular}
[c]{cccccc}\hline\hline
Variable name & Obs. & Mean & Std. Dev. & Min & Max\\\hline
\multicolumn{1}{l}{Reported profit $\Pi_{{\small N}}$ \ (thousand \$)} &
1,503 & \multicolumn{1}{r}{-56.0} & \multicolumn{1}{r}{731.7} &
\multicolumn{1}{r}{-22,283} & \multicolumn{1}{r}{1,764.4}\\
\multicolumn{1}{l}{Unreported profit $\Delta\Pi(\Gamma)$ (thousand \$)} &
1,323 & \multicolumn{1}{r}{423.3} & \multicolumn{1}{r}{2,705.1} &
\multicolumn{1}{r}{0} & \multicolumn{1}{r}{47,734.2}\\
\multicolumn{1}{l}{Required papers $All\_papers$ (number)} & 1,503 &
\multicolumn{1}{r}{8.1} & \multicolumn{1}{r}{3.027} & \multicolumn{1}{r}{1} &
\multicolumn{1}{r}{30}\\
\multicolumn{1}{l}{Capital $M$ (thousand \$)} & 1,473 &
\multicolumn{1}{r}{554.3} & \multicolumn{1}{r}{4,062.1} &
\multicolumn{1}{r}{0.066} & \multicolumn{1}{r}{63,645.6}\\
\multicolumn{1}{l}{Workers (number)} & 1,503 & \multicolumn{1}{r}{42.7} &
\multicolumn{1}{r}{83.9} & \multicolumn{1}{r}{2} & \multicolumn{1}{r}{1,040}\\
\multicolumn{1}{l}{Official paperwork processing cost $F_{{\small L}}$ (\$)} &
1,483 & \multicolumn{1}{r}{40,981} & \multicolumn{1}{r}{321,482} &
\multicolumn{1}{r}{10.6} & \multicolumn{1}{r}{9,361,830}\\
\multicolumn{1}{l}{Actual paperwork processing cost $F$ (\$)} & 1,486 &
\multicolumn{1}{r}{42,950} & \multicolumn{1}{r}{323,543} &
\multicolumn{1}{r}{21.2} & \multicolumn{1}{r}{9,361,830}\\
\multicolumn{1}{l}{Aggregate bribe $Bribe$ (\$)} & 1,483 &
\multicolumn{1}{r}{1,994} & \multicolumn{1}{r}{27,000.5} &
\multicolumn{1}{r}{0} & \multicolumn{1}{r}{800,000}\\
\multicolumn{1}{l}{Reported aggregate bribe $Rep\_bribe$ (\$)} & 1,347 &
\multicolumn{1}{r}{2,195} & \multicolumn{1}{r}{28,323.5} &
\multicolumn{1}{r}{0} & \multicolumn{1}{r}{800,000}\\
\multicolumn{1}{l}{Disputes $1_{disp}$ ($1$, if$\ \#$ disputes $\geq1$)} &
\multicolumn{1}{r}{1,503} & \multicolumn{1}{r}{0.220} &
\multicolumn{1}{r}{0.415} & \multicolumn{1}{r}{0} & \multicolumn{1}{r}{1}\\
\multicolumn{1}{l}{Official paperwork process. time $t_{{\small min}}$ (days)}
& 1,497 & \multicolumn{1}{r}{11.6} & \multicolumn{1}{r}{11.1} &
\multicolumn{1}{r}{0} & \multicolumn{1}{r}{150}\\
\multicolumn{1}{l}{Actual paperwork processing time $t$ (days)} & 1,497 &
\multicolumn{1}{r}{16.1} & \multicolumn{1}{r}{14.7} & \multicolumn{1}{r}{0} &
\multicolumn{1}{r}{170}\\
\multicolumn{1}{l}{Bureaucratic delay $Delay$ (days)} & 1,497 &
\multicolumn{1}{r}{4.5} & \multicolumn{1}{r}{6.5} & \multicolumn{1}{r}{0} &
\multicolumn{1}{r}{100}\\\hline\hline

\end{tabular} \label{table:data_summary}
 \caption*{\footnotesize{Note: Details on the construction and measurement of variables are in the online appendix.}}
}
\end{table}
\normalsize

\subsection{Estimation approach}\label{subsec:estimation_approach}

Because the aggregate bribe in \ref{eq:EM1} linearly depends on variables of interest and since we have three observations for each firm in the data, we use the panel linear regression model, with each panel containing data from 2012-2014 for each firm. We use the within estimator with firm fixed effects: we subtract the panel-specific means from observations in each panel and then apply OLS to the demeaned data. We calculate heteroskedasticity robust standard errors that are clustered at the firm level.

Because we eliminate all firm-specific means from observations, our estimates rely only on
the cross-time and within-firm variation in the data. Consequently, our estimates are unbiased with respect to all time-invariable independent variables. These variables include firms' bureaucratic connections, political affiliations, prior corruption experience, the level of compliance with paperwork requirements, and the level of transparency. Further, because there were no bureaucratic reforms in Tajikistan in 2012-2014, these variables include location-specific characteristics and industry-wide shocks: the degree of enforcement of laws, general levels of corruption, bureaucratic congestion rates, and industry-wide productivity shocks. 

Lastly, the variables with excluded effects are the unobservable sample selection variables and interviewer-specific measurement errors. The effects of these variables are excluded because the selection of firms into the sample and the recording of responses by interviewers relied on the ability of interviewers to convince firm managers of confidentiality of disclosed data, since firm managers had to directly admit to illegal activities such as bribe payments and hidden profits. Because the same interviewers collected data from the same firms in all years, we can model the sample selection equation and the interviewer-specific measurement error as a function of the indicator variables of individual interviewers. As individual interviewer indicator variables remain constant in each panel, the process of demeaning eliminates the effect of these sample selection and measurement variables.

The summary statistics in Table \ref{table:data_summary} show that the data are skewed. Some smaller firms in the data report paying bribes as small as US \$1 and some larger firms report paying bribes as large as US \$ 800,000. Similar patterns hold for firms' employment
levels, capital size, and bureaucratic burden. To reduce skewness in the data,
we normalize all variables in \ref{eq:EM1} by the total number of workers
($Workers$). This normalization reduces the standard errors of the controls by several
orders of magnitude\footnote{See Table C7 in the online appendix for the 95\%
confidence intervals and standard errors of normalized variables.}.

The model of corruption in Section \ref{sec:model} does not account for all bribery
mechanisms and ignores customary transfers that firms
make to bureaucrats. These customary transfers reflect cultural norms in our
market for bribes, and represent socially accepted gifts or donations
from firms to public servants.  Firms that have more paperwork requirements make larger customary
transfers, as acquisition of each certificate is accompanied by transfer of a socially accepted gift to a bureaucrat responsible for issuing the
certificate. To account for customary transfers, we use firms' reports of the total number of required legal papers ($All\_papers$) outside and during disputes ($All\_papers1_{disp}$) and the
intercept. This results in adding the terms $\frac{All\_papers}{Workers}$ and $\frac{All\_papers1_{disp}}{Workers}$ with coefficients $\beta_{AP}$ and $\beta_{APD}$, respectively, when constructing the empirical specification based on \ref{eq:EM1}. Lastly, we use year fixed effects ($1_{Year=2013}$ and $1_{Year=2014}$) to account for economy-wide
shocks. With these adjustments, we obtain the following empirical
specification \ref{eq:EM11}:

\small{
\begin{align}\label{eq:EM11}
\frac{\text{{}}Bribe}{Workers}  &  =\beta_{{\small \Pi}_{N}}\frac{\Pi_{N}%
}{Workers}+(1_{NS})\frac{\Delta\Pi(\Gamma=0)1_{P}1_{disp}}{Workers}%
+(1_{SwC}\lambda)\frac{\Delta\Pi(\Gamma=1)1_{P}}{Workers} \nonumber\\
&  +\beta_{{\small M}}\frac{M1_{disp}}{Workers}+\beta_{{\small F}_{L}}%
\frac{F_{L}1_{disp}}{Workers}+\beta_{{\small t}_{min}}\frac{t_{min}1_{disp}%
}{Workers}+(-s_{{\small N}}1_{B})\frac{Delay(1-1_{P})1_{disp}}{Workers}%
\nonumber\\
&  +(-s_{{\small P}})\frac{Delay1_{P}1_{disp}}{Workers}+(1_{SwC}g)\frac
{Delay1_{P}}{Workers}+\beta_{2013}1_{(Year=2013)}+\beta_{2014}1_{(Year=2014)}\nonumber\\
&  +\beta_{AP}\frac{All\_papers}{Workers}+\beta_{APD}\frac{All\_papers1_{disp}%
}{Workers}+Intercept \tag{EM1.1}
\end{align}
}
\normalsize

\subsection{Estimation results}
Estimates of specification \ref{eq:EM11} are reported in the first column of
Table \ref{table:coefs}. We rely on sign restrictions in Table \ref{table:ident_inf_scen} to interpret the empirical evidence in Table \ref{table:coefs}. 

\paragraph*{Evidence of coexistence of both bribery types:} Recall from the discussion of identification in Section \ref{subsec:identification} that $1_{SwC}\lambda > 0$ and $1_{SwC} g >0$ identify the presence of non-extortionary bribery. Here, both estimates are indeed positive and statistically significant.  The estimate of $1_{SwC}\lambda$ is $0.039$, indicating that the $P$-type firm's non-extortionary bribe is about 4\% of a thousand USD in firms' actual profits. The estimate of $1_{SwC}g$ is $8.929$, indicating that the $P$-type firm's daily participation cost in a bribe contest is about US \$ 9.

\small{
\begin{table}[hbt!]
\caption{Estimated coefficients.}
\small{
\begin{tabular}
[c]{cllllll}\hline\hline
Coefficients & \multicolumn{2}{c}{\ref{eq:EM11}} & \multicolumn{2}{c}{EM1.2} &
\multicolumn{2}{c}{EM1.3}\\\hline
$\beta_{\Pi_{N}}$ & 5.029** & (2.238) & 6.159*** & (1.593) & 5.014** &
(2.242)\\
$1_{NS}$ & -0.007 & (0.098) & 0.017 & (0.118) &  & \\
$1_{SwC}\lambda$ & 0.039*** & (0.007) & 0.039*** & (0.008) &  & \\
$\beta_{M}$ & -0.022 & (0.046) & -0.026 & (0.046) & -0.026 & (0.049)\\
$\beta_{F_{L}}$ & -0.002 & (0.005) & -0.003 & (0.004) & -0.003 & (0.005)\\
$\beta_{t_{min}}$ & 1.729 & (3.268) & 2.557 & (4.055) & 1.873 & (3.334)\\
$-s_{N}1_{B}$ & -5.595 & (21.625) & -2.090 & (21.968) & -6.992 & (21.769)\\
$-s_{P}$ & -8.804** & (4.116) & -9.142** & (4.650) & -9.418** & (4.178)\\
$1_{SwC}g$ & 8.929** & (4.143) & 9.539** & (4.849) & 9.578** & (4.179)\\
$\beta_{AP}$ & 27.626** & (13.120) & 30.258** & (14.438) & 24.576* &
(14.672)\\
$\beta_{APD}$ & 14.693 & (9.133) & 12.725 & (9.662) & 15.627 & (9.590)\\
$Intercept$ & 35.930*** & (7.971) & 39.397*** & (8.727) & 37.916*** &
(8.816)\\
Year FX & \multicolumn{2}{c}{Yes} & \multicolumn{2}{c}{Yes} &
\multicolumn{2}{c}{Yes}\\
Firm FX & \multicolumn{2}{c}{Yes} & \multicolumn{2}{c}{Yes} &
\multicolumn{2}{c}{Yes}\\
Within $R^{2}$ & \multicolumn{2}{c}{0.087} & \multicolumn{2}{c}{0.098} &
\multicolumn{2}{c}{0.084}\\
Between $R^{2}$ & \multicolumn{2}{c}{0.633} & \multicolumn{2}{c}{0.664} &
\multicolumn{2}{c}{0.651}\\
Overall $R^{2}$ & \multicolumn{2}{c}{0.485} & \multicolumn{2}{c}{0.522} &
\multicolumn{2}{c}{0.499}\\
Observations & \multicolumn{2}{c}{1,279} & \multicolumn{2}{c}{1,202} &
\multicolumn{2}{c}{1,279}\\\hline\hline
\end{tabular} \label{table:coefs}
}
\caption*{\footnotesize{(a) The top row lists empirical specifications;\\ (b) The dependent variable is $\frac{Bribe}{workers}$ in
columns 1,3 and $\frac{Rep\_Bribe}{workers}$ in column 2;\\ (c) Robust standard errors in parentheses are clustered by
firm id's; \\ (d) $\Pi_{N}$, $\Delta\Pi$, and $M$ are measured in
thousand US \$; \\ (e) ***,**,* - statistical significance at 1\%, 5\%, 10\%
level.}}
\end{table}
}
\normalsize

Recall from Section \ref{subsec:identification} that $s_P>0$ and $s_N 1_B \geq 0$ give evidence of extortionary bribery. The estimate of $-s_{P}$ is $-8.804$, and it is statistically significant, suggesting that the daily cost of time
of the $P$-type firm is about US \$ 9. The statistically significant negative
relationship between bribes and delay during bureaucratic disputes, measured by the estimate of $-s_{P}$, indicates that a day of bureaucratic delay reduces extortionary bribe by about US \$ 9. Note that estimates of $g$ and $s_{P}$ are similar in size, which shows that the $P$-type firm has identical daily participation and time
costs. This is not surprising, since $g$ and $s_{P}$ essentially measure the
same parameter: managers' daily cost of time when negotiating with corrupt bureaucrats either to gain access to government resources during bribe contests or to reduce
bribe demands during bureaucratic disputes. The magnitudes of the estimated
$s_{P}$ and $g$ support this argument, as both are similar in size to an average daily
salary of a firm manager (1.5 of an average daily salary of a worker).

The estimate of $-s_{N}1_{B}$ is negative but statistically insignificant. We
interpret this result as evidence that the
$N$-type firm cannot afford to pay any extortionary bribe ($1_{B}=0$). Alternatively, it can be interpreted as the evidence of the negligent cost of time ($s_{N}=0$) without assuming away the $N$-type
firm's ability to pay a positive extortionary bribe ($1_{B}=1$). However, note that the $N$-type firm with no cost of time will always prefer to accept $B2$'s offer with the longest waiting time and zero extortionary bribe. Hence, whether the $N$-type firm has positive cost of time and is unable to pay any extortionary bribe ($s_N>0$ and $1_{B}=0$) or has zero cost of time
and is able to pay a positive extortionary bribe ($s_N=0$ and $1_{B}=1$), it never pays
an extortionary bribe in equilibrium. Since $s_{N}$ is not identified when
$1_{B}=0$, we cannot make a claim about the $N$-type firm's daily cost of time ($s_{N}$).

\paragraph*{Evidence for the SwC informational scenario:}  Recall from Section \ref{subsec:identification} and Table \ref{table:ident_inf_scen} that the parameter values $1_{SwC}\lambda > 0$, $1_{SwC} g >0$, and $1_{NS}=0$ identify the SwC scenario. Estimates in Table \ref{table:coefs} are consistent with these parameter values, indicating the empirical support for the SwC scenario. The
estimate of $1_{SwC}\lambda$ is $0.039$, and it indicates that the $P$-type firm's non-extortionary bribe is about 4\% of a thousand USD in actual profits. The estimate of $1_{SwC}g$ is $8.929$, indicating that the $P$-type
firm's daily participation cost in a bribe contest is approximately US \$ 9. Both estimates are statistically significant.

Additional supporting evidence for the SwC scenario comes from the statistically insignificant estimate of $1_{NS}$. This is because the empirical support for the SwC scenario simultaneously rejects the other two scenarios, as all three scenarios are mutually exclusive.   

\paragraph*{Other coefficients:} The estimate of $\beta_{{\small \Pi}_{N}}$ is positive and statistically significant: a thousand dollar increase in reported profit raises aggregate bribe by almost US \$ 5. Estimates of
coefficients of capital ($\beta_{M}$), official paperwork processing cost
($\beta_{F_{L}}$), and official paperwork processing time ($\beta_{t_{min}}$)
are statistically insignificant. The statistical insignificance of $\beta_{M}%
$, $\beta_{F_{L}}$, and $\beta_{t_{min}}$ is due to the use of the fixed
effects estimator that eliminates firm-specific means. Since capital ($M$), official paperwork processing cost
($F_{L}$), and official paperwork processing time ($t_{min}$) do not
sufficiently vary over time, their effects are completely eliminated.

Estimates of coefficients measuring customary transfers are positive and
statistically significant. The estimate of the coefficient of the total number
of required legal papers ($\beta_{AP}$) is 27.626. Its magnitude indicates
that an additional certificate is accompanied by a socially accepted gift to bureaucrats of US \$ 27. The coefficient of the total number of required legal papers during disputes
($\beta_{APD}$) is statistically insignificant, suggesting that customary transfers are not made during bureaucratic disputes. The estimate of the intercept is statistically significant, and its magnitude is an indication of a baseline customary annual transfer of US \$ 36 for firms with registration at the tax office (all firms in the data).

\subsection{Limitations and robustness checks}

There are two main caveats to our results. The first caveat concerns the non-causal relationship between aggregate bribes and bureaucratic delays in specifications \ref{eq:EM1} and \ref{eq:EM11}. Recall that specification \ref{eq:EM1} is derived under the assumption that the corrupt bureaucrat B2 maximizes the size of the extortionary bribe by reducing the paperwork processing time. We can derive an alternative specification from the same equilibrium conditions of the theoretical model, where B2 maximizes the paperwork processing time by reducing the size of the extortionary bribe demand. Hence, to interpret our estimation results as causal, we have to maintain the assumption that corrupt bureaucrats view bureaucratic delays as instruments in their objective to maximize extortionary bribe demands. 

The second caveat concerns the omitted variable bias. Although our estimation approach controls for unobservable firm-specific characteristics that do not vary in 2012-2014, our results may be biased due to omitted firm-specific characteristics that are jointly correlated with the dependent variable (aggregate bribes) and independent variables (firms' profits and bureaucratic delays) and that could vary in 2012-2014. To our knowledge, there were no productivity shocks and no changes in the bureaucratic environment that could jointly affect bribe levels, bureaucratic delays, and firms' profits in Tajikistan in 2012-2014. This claim is supported by the statistical insignificance of the demeaned official paperwork costs ($\beta_{F_{L}}$) and paperwork processing times ($\beta_{t_{min}}$) in the specification \ref{eq:EM11}. As an additional control for possible annual shocks to firms from changes in the bureaucratic environment or from productivity adjustments, we include controls for the year fixed effects. These year fixed effects in Table \ref{table:coefs} are statistically insignificant, suggesting that the omitted variables bias is unlikely to contaminate our estimates.

We implement robustness checks to test the stability of our results. In the specification EM1.2, we replace the indirectly measured bribe $\frac{\text{{}}Bribe}{Workers}$ with the directly measured bribe $\frac{\text{{}}Rep\_Bribe}{Workers}$. The estimates of the specification EM1.2 are reported in the second column of Table \ref{table:coefs}. These estimates are of similar size and statistical significance to those of \ref{eq:EM11}, with the estimated coefficients of $\beta_{\Pi_{N}}$, $g$, $s_{P}$, and $\beta_{AP}$ slightly larger than those of \ref{eq:EM11}. This suggests that the measurement error is unlikely to contaminate our estimates, as specifications with directly and indirectly measured aggregate bribes give similar results.

In the specification EM1.3, we omit controls for unreported profits $\frac{\Delta\Pi(\Gamma=0)1_{disp}}{Workers}$ and $\frac{\Delta\Pi(\Gamma=1)}{Workers}$, because the measures for unreported profits can be interpreted as perceived rather than actual, and the statistical significance of the delay coefficients in specification \ref{eq:EM11} could be driven by the collinearity of bureaucratic delays and unreported profits. The estimates of EM1.3 are reported in the third column of Table \ref{table:coefs}. The estimated delay coefficients are similar to those of \ref{eq:EM11}, which implies that the statistical significance of delay coefficients is not affected by the variation in unreported profits. Note that without controls for unreported profits, it is impossible to estimate $1_{NS}$ and $\,1_{SwC}\lambda$ and, therefore, identify a specific informational scenario.

We extend the baseline specification \ref{eq:EM11} to include three types of firms: firms with non-negative, medium, and large profits. We discuss this extension in the online appendix. The estimates of this specification are consistent with those of \ref{eq:EM11} but are less precise. The lower precision of the estimates is due to the fewer observations available for estimation of parameters for each firm type.

\section{Discussion and policy recommendations}\label{sec:discussion}

In this section, we discuss estimation results and provide policy recommendations. 

\textit{The ``grease vs. sand" bribery debate:} We find that firms simultaneously pay both types of
bribes: extortionary and non-extortionary. The empirical evidence for the extortionary
bribery in the data supports the ``sand in the wheels" bribery hypothesis,
while the empirical evidence for the non-extortionary bribery supports the
``grease the wheels" bribery hypothesis. Hence, our empirical results indicate that both bribery hypotheses can simultaneously hold in the same market for bribes. 

Because of the coexistence of bribery types, the question of whether bribery benefits or harms individual
firms does not have an unequivocal answer in our market for bribes. On the one hand, the non-extortionary bribery can help individual firms by giving access to government contracts. On the other hand, extortionary bribery
hurts individual firms by reducing their profits and by subjecting all firms to the costly bureaucratic delay. Specifically, for
$N$-type firms, which can face only extortionary bribe demands, the net effect of bribery is
clearly negative. For $P$-type firms, which participate in both types of bribery,
the net effect can be positive or negative.

\textit{Aggregate bribes and delays:} How to reconcile our estimates in Table \ref{table:coefs} with
the aggregate patterns in the market for bribes? Recall from Figure \ref{fig:bribe_delay}B
that firms with positive profits face longer aggregate bureaucratic
delays than firms with non-positive profits. This seems to contradict our result that the positive-profit firms should face shorter bureaucratic delays because they choose to pay larger bribes. Note, however, that aggregate delays combine
participation times in bribe contests (that only the positive-profit firms enter) and
bureaucratic delays due to extortion, suggesting that aggregate bureaucratic delays can indeed be longer for positive-profit firms. Further, since both types of firms can make positive customary transfers, the aggregate bribes of both $N$-type and $P$-type can be positive.  

\textit{Policy recommendations:} Because we find that firms engage in extortionary and non-extortionary bribery in the same market for bribes, there are at least two sets of policies that can help to reduce the overall incidence of bribery. For firms involved in the non-extortionary bribery, the literature on corruption has long identified policies that can reduce the incidence of bribe transfers (see, for example, \cite{Banerjee2023}, \cite{Abbink2012}, \cite{Abbink2002}). These policies include rotations of bureaucrats across offices (\cite{abbink2004}), top-down and bottom-up monitoring of bureaucrats' and firms' actions (\cite{Abbink2006}, \cite{Serra2012}, \cite{RyvkinSerraTremewan2017}), introduction of competition among bureaucrats by allowing multiple bureaucrats to provide identical services (\cite{Drugov2010}), elimination of intermediaries that facilitate the matching of firms willing to pay bribes and bureaucrats willing to accept bribes (\cite{Drugov2014}).

For firms involved in extortionary bribery, the set of policy recommendations is different. Policies that restrict bureaucrats' ability to use differentiated paperwork processing times can limit corrupt bureaucrats' ability to extract larger bribes from more profitable firms. In particular, if bureaucrats are required to provide their services to all firms in the same amount of time, then no sorting of firms by profit level is possible, and all firms have to pay the same minimal extortionary bribe.

Further, because the size of the extortionary bribe negatively depends on the firm's outside option, any policies that raise outside options of firms facing extortionary bribe demands weaken corrupt bureaucrats' bargaining position and lead to smaller extortionary bribes. Hence,  holding everything else constant, lower fines for firms paying extortionary bribes, weaker punishment for firms that operate without licenses due to extortionary bribe demands, and facilitation of the ability of firms to shut down or restart operations should reduce extortionary bribery. 

The recommendation of a weaker punishment for firms involved in the extortionary bribery is consistent with those of \cite{Basu2011}, \cite{Abbink2014}, and \cite{Basu2016}, who argue that for a class of extortionary or harassment bribes, the punishment imposed on the demander of a bribe (corrupt bureaucrat) must be much more severe than the punishment imposed on the supplier of a bribe (firm).

\section{Conclusion}\label{sec:conclusion}

The market for bribes in our data is characterized by bribe payments that are consistent both with the ``grease" and the ``sand" bribery hypotheses. The evidence of a positive correlation between bureaucratic delays and bribes outside of bureaucratic disputes for positive-profit firms indicates that firms with financial resources willingly participate in contests for public resources that involve voluntary bribe payments. The evidence of a negative correlation between bureaucratic delays and bribes during bureaucratic disputes for positive-profit firms indicates that firms with financial resources prefer to pay larger bribes to reduce red tape that uninformed corrupt bureaucrats intentionally create to sort out firms that have financial resources. Uninformed corrupt bureaucrats are compelled to exploit bureaucratic delays because firms differ in their costs of time and because firms hide large portions of their actual profits. Our results provide a test of the presence of multiple types of bribery in the absence of detailed information about types of individual bribe payments. 

We do not find evidence of communication among corrupt bureaucrats in the data. Corrupt bureaucrats that receive non-extortionary bribes do not transmit relevant information about resources of bribe-payers to other corrupt bureaucrats. The lack of evidence of the transmission of bribery-relevant information suggests that the market for bribes in our paper is more accurately described as decentralized.

\small
\bibliographystyle{aer}

\bibliography{bib}

\section*{Online Appendix for \\ \textit{{Bribery, Secrecy, and Communication: Theory and Evidence from Firms}}}

\section*{A. Proofs of Lemmas}
\subsection*{Proof of Lemma 2} \label{app:pf_lem_SC}
\begin{proof}
Since $U_{P,\Gamma=1}>U_{P,\Gamma=0}>U_{N,\Gamma}$, the participation
constraint $IR_{P,\Gamma=0}$ never binds and can be ignored. Further, $IC_{N}$
and $IC_{P,\Gamma=0}$ together imply that $t_{N}\geq t_{P,\Gamma=0}$, which
means that $IC_{N}$ can be replaced with monotonicity condition $t_{N}\geq
t_{P,\Gamma=0}$. In particular, by adding $IC_{N}$ and $IC_{P,\Gamma=0}$, we
get the condition $(s_{P}-s_{N})(t_{N}-t_{P,\Gamma=0})\geq0$, which holds\ as
long as $t_{N}\geq t_{P,\Gamma=0}$. Hence, we have only two constraints,
$IR_{N}$ and $IC_{P,\Gamma=0}$, and the monotonicity condition $t_{N}\geq
t_{P,\Gamma=0}$.

Next, note that $B2$ can strictly improve his payoff by increasing $t_{N}$
until $IR_{N}$ is binding without violating the monotonicity condition and
$IC_{P,\Gamma=0}$. The increase in $t_{N}$ does not violate $IC_{P,\Gamma=0}$,
because it lowers only the right-hand side of $IC_{P,\Gamma=0}$. Given the
binding $IR_{N}$, $B2$'s processing time offer for the $N$-type firm is
$t_{N}^{SC}=\frac{\Pi_{N}-M-F_{L}}{s_{N}}$. In addition, $B2$ can strictly
improve his payoff by raising $F_{P,\Gamma=0}$ to make $IC_{P,\Gamma=0}$
binding without violating the monotonicity condition and $IR_{N}$.

$B2$'s optimization problem can be simplified by expressing $F_{P,\Gamma=0}$
in terms of $t_{P,\Gamma=0}$ given the binding $IC_{P,\Gamma=0}$. Then, $B2$'s
problem is to choose $t_{P,\Gamma=0}$ to maximize $W-\frac{\theta}%
{\theta+(1-\theta)\phi}\frac{cs_{N}}{\Pi_{N}-M-F_{L}}+\frac{(1-\theta)\phi
}{\theta+(1-\theta)\phi}\left(s_{P}\frac{\Pi_{N}-M-F_{L}}{s_{N}}-s_{P}%
t_{P,\Gamma=0}-\frac{c}{t_{P,\Gamma=0}}\right)$. The optimization yields the
processing time for the $P$-type firm of $t_{P,\Gamma=0}^{SC}=\sqrt{\frac
{c}{s_{P}}}$. The optimal processing times for both firm types, $t_{N}%
^{SC}=\frac{\Pi_{N}-M-F_{L}}{s_{N}}$ and $t_{P,\Gamma=0}^{SC}=\sqrt{\frac
{c}{s_{P}}}$, satisfy the monotonicity condition ($t_{P,\Gamma=0}^{SC}%
=\sqrt{\frac{c}{s_{P}}}<\sqrt{\frac{c}{s_{N}}}<\frac{\Pi_{N}-M-F_{L}}{s_{N}%
}=t_{N}^{SC}$), implying that $IC_{N,\Gamma=0}$ is also satisfied. Lastly, we
can find the $P$-type firm's certificate fee, $F_{P,\Gamma=0}^{SC}$, by
plugging in optimal processing times $t_{N}^{SC}=\frac{\Pi_{N}-M-F_{L}}{s_{N}%
}$ and $t_{P,\Gamma=0}^{SC}=\sqrt{\frac{c}{s_{P}}}$ into the binding
$IC_{P,t_{P,\Gamma=0}}$.
\end{proof}

\subsection*{Proof of Lemma 4} \label{ap:pf_lem_SwC}
\begin{proof}
Since $U_{P,\Gamma=1}>U_{P,\Gamma=0}>U_{N,\Gamma}$, we can discard $IR$
constraints $IR_{P,\Gamma=0}$ and $IR_{P,\Gamma=1}$, as they always hold.
$IC_{N}(P,\Gamma=0)$ and $IC_{P,\Gamma=0}(N)$ together imply that $t_{N}\geq
t_{P,\Gamma=0}$, which means that we can replace $IC_{N}(P,\Gamma=0)$ with
monotonicity condition $t_{N}\geq t_{P,\Gamma=0}$. Similarly, $IC_{N}%
(P,\Gamma=1)$ and $IC_{P,\Gamma=1}(N)$ together imply that $t_{N}\geq
t_{P,\Gamma=1}$, which means that we can replace $IC_{N}(P,\Gamma=1)$ with
monotonicity condition $t_{N}\geq t_{P,\Gamma=1}$. Next, $IC_{P,\Gamma=0}(N)$
and $IC_{P,\Gamma=1}(P,\Gamma=0)$ are equivalent to $IC_{N}(P,\Gamma=1)$,
which means that $IC_{N}(P,\Gamma=1)$ can be ignored. Lastly, $IC_{P,\Gamma
=0}(P,\Gamma=1)$ and $IC_{P,\Gamma=1}(P,\Gamma=0)$ together imply the equality%

\[
s_{P}(t_{P,\Gamma=0}-t_{P,\Gamma=1})=F_{P,\Gamma=1}-F_{P,\Gamma=0}.
\]

Hence, $B2$'s constraints reduce to
\begin{gather*}
IR_{N}:\Pi_{N}-s_{N}t_{N}-F_{L}\geq M\text{,}\\
IC_{P,\Gamma=0}(N):s_{P}(t_{N}-t_{P,\Gamma=0})\geq F_{P,\Gamma=0}%
-F_{L}\text{,}\\
s_{P}(t_{P,\Gamma=0}-t_{P,\Gamma=1})=F_{P,\Gamma=1}-F_{P,\Gamma=0}\text{,}\\
t_{N}\geq t_{P,\Gamma=0}\text{, }\\
t_{N}\geq t_{P,\Gamma=1}\text{.}%
\end{gather*}

$B2$ can raise his expected payoff by increasing $t_{N}$ until $IR_{N}$ binds
without violating any of the remaining constraints. This gives the optimal
time for the $N$-type firm: $t_{N}^{SwC}=\frac{\Pi_{N}-M-F_{L}}{s_{N}}$.
Similarly, $B2$ can raise his payoff by simultaneously raising $t_{P,\Gamma
=0}$ and $t_{P,\Gamma=1}$ until $IC_{P,\Gamma=0}(N)$ binds without violating
any of the remaining constraints. This yields the binding $IC_{P,\Gamma=0}(N)$:
\begin{gather*}
s_{P}(t_{N}-t_{P,\Gamma=0})=F_{P,\Gamma=0}-F_{L}\\
\iff \text{ }F_{P,\Gamma=0}=s_{P}(t_{N}-t_{P,\Gamma=0})+F_{L\text{.}}%
\end{gather*}

Lastly, the equality constraint $s_{P}(t_{P,\Gamma=0}-t_{P,\Gamma
=1})=F_{P,\Gamma=1}-F_{P,\Gamma=0}$ can be written as follows:
\begin{align*}
F_{P,\Gamma=1}  &  =s_{P}(t_{P,\Gamma=0}-t_{P,\Gamma=1})+F_{P,\Gamma=0}\\
&  =s_{P}(t_{P,\Gamma=0}-t_{P,\Gamma=1})+s_{P}(t_{N}-t_{P,\Gamma=0})+F_{L}\\
&  =s_{P}(t_{N}-t_{P,\Gamma=1})+F_{L}\text{.}%
\end{align*}

By replacing $t_{N}$ with $t_{N}^{SwC}$, $F_{P,\Gamma=0}$ with $s_{P}%
(t_{N}-t_{P,\Gamma=0})+F_{L}$, and $F_{P,\Gamma=1}$ with $s_{P}(t_{N}%
-t_{P,\Gamma=1})+F_{L}$ in $B2$'s objective function, we obtain%

\[
W-\theta\frac{c}{t_{N}^{SwC}}+\phi(1-\theta)(s_{P}(t_{N}^{SwC}-t_{P,\Gamma
=0})-\frac{c}{t_{P,\Gamma=0}})+(1-\phi)(1-\theta)(s_{P}(t_{N}^{SwC}%
-t_{P,\Gamma=1})-\frac{c}{t_{P,\Gamma=1}})\text{. }%
\]

The maximization of this objective function with respect to $t_{P,\Gamma=0}$
and $t_{P,\Gamma=1}$ gives identical processing times for the $P$-type firm
with and without the contract, $t_{P}^{SwC}=t_{P,\Gamma=0}^{SwC}=t_{P,\Gamma
=1}^{SwC}=$ $\sqrt{\frac{c}{s_{P}}}$. The optimal processing times
$t_{P,\Gamma=0}^{SwC}$ and $t_{P,\Gamma=1}^{SwC}$ satisfy monotonicity
conditions $t_{N}\geq t_{P,\Gamma=0}$ and $t_{N}\geq t_{P,\Gamma=1}$ and

\[
t_{N}^{SwC}=\frac{\Pi_{N}-M-F_{L}}{s_{N}}>\sqrt{\frac{c}{s_{N}}}>\sqrt{\frac
{c}{s_{P}}}=t_{P,\Gamma=0}^{SwC}=t_{P,\Gamma=1}^{SwC}\text{.}%
\]

By plugging optimal processing times into $F_{P,\Gamma=0}$ and $F_{P,\Gamma
=1}$, we obtain optimal fees for the $P$-type firm with and without the
contract: $F_{P}^{SwC}=F_{P,\Gamma=1}^{SwC}=F_{P,\Gamma=0}^{SwC}=s_{P}(t_{N}%
^{SwC}-t_{P}^{SwC})$ $+F_{L}$.

Since extortionary bribe $EB_{P}^{SwC}=F_{P}^{SwC}-F_{L}$ and delay $m_{P}%
^{SwC}=t_{P}^{SwC}-t_{min}$ of the $P$-type firm do not depend on participation
in the bribe contest in period $1$, the $P$-type firm's payoff from entering
the contest strictly exceeds that from not entering. Hence, it is dominant
strategy for the $P$-type firm to enter to contest in period $1$, and $\phi
=0$. 
\end{proof}



\section*{B. Extensions and additional estimation results}

\paragraph*{Uncertain acquisition of the contract in the bribe contest:} \label{ext:unc}
\normalfont
Let $\mu$ denote the probability of the $P$-type firm winning in the bribe contest upon entry. In this extension, suppose that $\mu<1$. 
Recall that when $\mu=1$ (baseline model), the $P$-type firm never enters the bribe contest in the NS and SC scenarios. The same must hold true for $\mu<1$, and the aggregate bribe of the $P$-type firm is unchanged for any $\mu\leq1$. 

In the SwC scenario, the $P$-type firm enters the bribe contest in equilibrium, but its extortionary bribe does not depend on its actual profit or participation in the bribe contest. Hence, the $P$-type firm's extortionary bribe is unchanged for any $\mu\leq1$. 

The $P$-type firm's non-extortionary bribe cost in the SwC scenario depends on whether the firm is the
winner in the bribe contest, which happens with probability $\mu$. In this scenario, the $P$-type firm's non-extortionary bribe cost is $NBC_{P}(\Gamma=1)=$ $gn+\lambda\Pi_{P}(\Gamma=1)$ with probability $\mu$ and $NBC_{P}(\Gamma=0)=$ $gn$ with probability $1-\mu$. The $P$-type firm enters the bribe contest only if its
expected payoff from entry $\mu(1-\kappa)V$ in period 1 exceeds its cost $gn+\lambda\mu
\Pi_{P}(\Gamma=1)$ in period 1, or if the probability of winning in the bribe contest upon entry is
sufficiently large, $\mu>\mu^*=\frac{gn}{(1-\kappa)V-\lambda\Pi_{P}(\Gamma=1)}$.

To incorporate the case of $\mu<1$ into the model, we introduce the indicator of the $P$-type firm winning in the bribe contest, $1_{W}=1$. Then, for any $\mu\in(\mu^*,1]$, the equilibrium aggregate bribe of the $P$-type firm in the SwC scenario is $Bribe_{P}^{SwC} =EB_{P}^{SwC}+1_{W}NBC_{P}(\Gamma=1)+(1-1_{W})NBC_{P}(\Gamma=0)$. In this expression, $EB_P^{SwC}=EB_P^{SC}$, $NBC_P(\Gamma=1)=gn+\lambda\Pi_{P}(\Gamma=1)$, and $NBC_P(\Gamma=0)=gn$ (sunk cost of contest participation). This gives
\begin{equation}\label{eq:eq_bribe_P_SwC_mu}
Bribe_{P}^{SwC}  =\frac{s_{P}}{s_{N}} \left(\Pi_{N}- M-
F_{L} \right)-s_{P}t_{min}-s_{P}m_{P}^{SwC}+gn+1_{W}\lambda\Pi_{P}(\Gamma=1)
\end{equation}
We use this specification for the aggregate bribe in the SwC scenario in all extensions in the online appendix. 

\paragraph*{Extension to multiple firm types:}
\normalfont With more than two (or a continuum) of firm types, the results in all three Propositions of the main text remain unchanged: the $N$-type firm pays no bribe and faces the longest delay,
the highest-profit firm with the highest cost of time pays the largest bribe
and faces the shortest delay, and firms with intermediate profits and time costs may pay smaller bribes and have longer delay than the highest-profit firm. 

The monotonicity of bribes by profit level (more
profitable firms pay larger bribes) holds if the $\frac{\partial}{\partial
s}(\frac{\partial U}{\partial F}/\frac{\partial U}{\partial t})<0$ (Spence-Mirrlees condition), which is
satisfied for the firm's payoff function in this paper. Since the strict ordering of
bribes by profit level (no pooling of types) holds for any distribution
function of types satisfying the monotone hazard rate property, whether the bribe of a firm with intermediate profit is strictly less than that of the firm with the highest profit depends on the particular distribution function of firms by profit level.

Let the $N$-type firm represent a non-positive-profit firm, the $MD$-type firm represent a
firm with positive profit below the median, and the $H$-type firm represent a firm
with positive profit above the median. As in the case with two types, the $N$-type firm does not pay
any non-extortionary bribe by avoiding participation in the bribe contest and
does not pay any extortionary bribe by choosing maximum bureaucratic delay.
The $MD$-type firm with the medium profit and the $H$-type firm with the large profit enter
the bribe contest in period 1, and therefore incur positive daily
participation costs $g_{MD}$ and $g_{H}$, respectively. When facing an
extortionary bribe request in period 2, both the $MD$-type firm and the $H$-type
firm prefer to pay positive extortionary bribes and have shorter delays than the
$N$-type firm due to larger costs of time, $s_{MD}$ and $s_{H}$, respectively.
Lastly, the more profitable $H$-type firm has higher cost of time than the
$MD$-type firm, which gives the following ranking of time costs: $s_{H}%
>s_{MD}>s_{N}$. Note that actual extortionary bribes and delays of $MD$-type and
$H$-type firms may or may not differ. In a pooling equilibrium, both types of firms choose the same bribe-delay combination, and in a separating
equilibrium, the bribe-delay combinations of $MD$-type and $H$-type firms are different.

In Table B1, we report estimated coefficients of the baseline
specification EM1.1 of the main text with separately estimated participation costs ($g_{MD}$
and $g_{H}$) and costs of time ($s_{MD}$ and $s_{H}$) of firms with positive
adjusted profits below the median ($MD$-type firms) and above the median
($H$-type firms). Because we split positive-profit firms into two groups and the
estimation of coefficients for each firm type exploits fewer observations,
the resulting estimates are not precise. Nevertheless, estimates indicate that
both $MD$-type and $H$-type firms participate in bribe contests and pay
extortionary bribes during disputes. The estimate of $s_{MD}$ is US\$ 6.6 and
the estimate of $g_{MD}$ is US\$ 4.3. The estimate of $s_{H}$ is US\$ 13.5 and
the estimate of $g_{H}$ is US\$ 13.5. As before, there is no empirical evidence that
$N$-type firms pay either extortionary or non-extortionary bribes.

Estimates of the model with three firm types are consistent with main
predictions of the model of corruption with only two firm types. Whether we
have two or three firm types, firms with positive profits engage in both types
of bribery. Because bureaucratic delay is more costly for more profitable
firms, these firms prefer to pay larger extortionary bribes. 


\paragraph*{Extension with the legal status of firms: }
In the theoretical model in the paper, we do not model the firms' decision to hide profits.  We only assume that actual profits of some firms may be unobservable to corrupt bureaucrats (possibly, due to tax avoidance or due to avoidance to pay bribes) and test this assumption empirically.

Further, in our model, the firm's tax or legal status does not affect the
firm's equilibrium aggregate bribe payment, because we assume that corrupt
bureaucrats receiving bribe payments  do not treat bribes coming
from the firms that pay their taxes fully any differently than those coming from the firms that pay their
taxes partially. In period 1, the non-extortionary bribe paid to bureaucrat
$B1$ does not depend on the firm's tax compliance status, because the non-extortionary
bribe is determined as a fraction of the firm's actual available resources
(actual profits). In period 2, the extortionary bribe is the solution to
bureaucrat $B2$'s constrained optimization problem, where $B2$ maximizes a sum
of his wage (W) and extortionary bribe net of the cost of the paperwork
processing time, independently of the firm's tax compliance or legal status.

We can extend the model and assume that the firm has an option to pay taxes
fully (be compliant) or pay taxes partially (be partially compliant) in period $0$. Then, $B2$ in
period $2$ would be facing a firm with 6 possible types: the $N$-type firm
that fully or partially complies with tax regulations, the $P$-type firm
with the contract that fully or partially complies with tax regulations, and
the $P$-type firm without the contract that fully or partially complies with
tax regulations. Screening models with more than two types are often
intractable, and solutions to tractable screening problems require strong
assumptions\footnote{See, for example, discussion in Laffont, Jean-Jacques, and
David Martimort. "The theory of incentives: the principal-agent model." The
theory of incentives. Princeton University Press, 2009).}.

However, we can still state that in the $NS$ scenario with six firm
types, the participation constraints of the $N$-type and the $P$-type firms
with or without the tax compliance status are binding. The solutions in the $%
SC$ and $SwC$ scenarios depend on whether the Spence-Mirrlees condition is
satisfied and whether the monotone hazard ratio property of the distribution
of firm's types is satisfied. Assuming that the Spence-Mirrlees condition
holds (more profitable firms are willing to pay larger bribes), the $IR$
constraint of a firm with the lowest profit and the $IC$ constraint of a
firm with the highest profit bind. For firms with intermediate profits,
depending on what we assume about the distribution of firm types, we may
have a pooling equilibrium with firms of different types paying the same
bribe or a separating equilibrium with firms of different types paying
different bribes. In general, the extension of the model in this direction
requires strong assumptions about (1) the ranking of profits of firms of
different types (with and without the contract and with and without the tax
compliance status) and (2) the density of firms of different types.

We can empirically test the consistency of our results with more than two types without deriving the equilibrium conditions. In our data, we have information on the legal status of firms. The legal
status indicator ($1_{legal}$) takes the value of $1$ for a firm that has all legal papers on hand and the value of $0$ for a firm with at least one
missing legal paper. We can use this indicator as a proxy for firms that
under-pay taxes, assuming that firms that avoid paying taxes also miss some legal papers, such as approvals from tax office,
licenses, or permits. As a reduced-form test, we estimate the baseline specification for firms that have full legal status and for firms that do not have full legal status by interacting profit and delay variables with the legal status indicator ($1_{legal}$). We report the estimation results in the Table B3 below.

The estimation results are consistent with the empirical results in the main text
and provide empirical support for the "Secrecy with Communication" scenario.
The estimates of $1_{NS}(1-1_{legal})$ and $1_{NS}1_{legal}$ are
statistically insignificant, while the estimate of $1_{SwC}1_{W}\lambda
(1-1_{legal})$ is statistically
significant. There is also no statistically significant effect of the legal
status on average bribe, as the coefficient of $1_{legal}$ is statistically
insignificant. Although the main results of the paper are unchanged, there are
differences in profit effects and costs of delay between firms with and
without the full legal status. For example, the extortionary bribe of the
P-type firm without the full legal status is about 4.6\% of the unreported
profit, with the estimate statistically significant at 1\% level. In
contrast, the extortionary bribe of the P-type firm with the full legal
status is about 18.7\% of the unreported profit, but this estimate is statistically insignificant. The estimate of
the same effect in the baseline model is 3.9\%, it is statistically
significant at 1\% level, and it is similar to that of the $P$-type firm
without the full legal status. Furthermore, the estimate of the daily cost of
delay of the $P$-type firm without the full legal status is US \$6
(estimates of $s_{P}(1-1_{legal})$ and $1_{SwC}g(1-1_{legal})$), while the
daily cost of delay of the $P$-type firm with the full legal status is US
\$5.5 (estimates of $s_{P}1_{legal}$ and $1_{SwC}g1_{legal}$). Both these
estimates are statistically insignificant. The estimate of the daily cost of
delay of the $P$-type firm in the baseline model is US \$8.9, and it is
statistically significant at 5\% level. Note that this estimate is bounded by the estimates for P-type firms with and without the full legal
status. Consistently with the estimates of the baseline model, the estimate of the daily cost of
delay of the $N$-type firm with or without the full legal status is statistically insignificant.

In general, while the estimates in this specification are consistent with
those of the baseline model, they are less precise. This is because we are
using fewer observations when estimating profit and delay coefficients
separately for firms with full and without full legal status. Hence, the
estimation results in the baseline specification should be interpreted as
average estimates for firms that pay taxes fully and for firms that pay
taxes partially.

\small{
\begin{align*}
&  \text{Table B1. Baseline specification with three firm types.}\\
&
\begin{tabular}
[c]{cllll}\hline\hline
Coefficients & \multicolumn{2}{c}{$\frac{Bribe}{workers}$} &
\multicolumn{2}{c}{$\frac{Rep\_Bribe}{workers}$}\\\hline
$\beta_{\Pi_{N}}$ & 5.030** & (2.239) & 6.161*** & (1.593)\\
$1_{CI}$ & -0.004 & (0.099) & 0.019 & (0.118)\\
$1_{CU}1_{W}\lambda$ & 0.039*** & (0.007) & 0.039*** & (0.008)\\
$\beta_{M}$ & -0.020 & (0.047) & -0.023 & (0.046)\\
$\beta_{F_{L}}$ & -0.002 & (0.005) & -0.003 & (0.004)\\
$\beta_{t_{min}}$ & 1.607 & (3.586) & 2.425 & (4.009)\\
$-s_{N}1_{B}$ & -6.603 & (21.710) & -3.044 & (22.078)\\
$-s_{MD}$ & -6.627 & (4.340) & -7.915* & (4,651)\\
$-s_{H}$ & -13.503* & (7.716) & -12.991 & (8.131)\\
$1_{CU}g_{MD}$ & 4.297 & (3.271) & 2.731 & (4.203)\\
$1_{CU}g_{H}$ & 13.584* & (7.711) & 13.269 & (8.111)\\
$\beta_{AP}$ & 26.785** & (12.911) & 29.623*** & (14.273)\\
$\beta_{APD}$ & 15.662* & (9.316) & 13.596 & (9.365)\\
$Intercept$ & 36.201*** & (7.908) & 39.899*** & (8.598)\\
Year FX & \multicolumn{2}{c}{Yes} & \multicolumn{2}{c}{Yes}\\
Firm FX & \multicolumn{2}{c}{Yes} & \multicolumn{2}{c}{Yes}\\
Within $R^{2}$ & \multicolumn{2}{c}{0.087} & \multicolumn{2}{c}{0.099}\\
Between $R^{2}$ & \multicolumn{2}{c}{0.633} & \multicolumn{2}{c}{0.666}\\
Overall $R^{2}$ & \multicolumn{2}{c}{0.484} & \multicolumn{2}{c}{0.523}\\
Observations & \multicolumn{2}{c}{1,279} & \multicolumn{2}{c}{1,202}%
\\\hline\hline
\end{tabular}
\\
&  \text{(a) The dependent variable is }\frac{Bribe}{workers}\text{ in column
1 and }\frac{Rep\_Bribe}{workers}\text{ in column 2;}\\
&  \text{(b) Robust standard errors in parentheses are clustered by firm
id's;}\\
&  \text{(c) }\Pi_{N}\text{, }\Delta\Pi\text{, and }M\text{ are measured in
thousand US \$;}\\
&  \text{(d) ***, **, * - statistical significance at 1\%, 5\%, 10\% level.}%
\end{align*}
}

\normalsize

\small{
\begin{align*}
&  \text{Table B2. Baseline specification EM1.1 with separately estimated }\\
&  \text{ intercepts and participation costs of }N\text{-type and }P\text{-type firms.}\\
&
\begin{tabular}
[c]{cllll}\hline\hline
Coefficients & \multicolumn{2}{c}{$\frac{Bribe}{workers}$} &
\multicolumn{2}{c}{$\frac{Rep\_Bribe}{workers}$}\\\hline
$\beta_{\Pi_{N}}$ & 5.095** & (2.204) & 6.171*** & (1.584)\\
$1_{NS}$ & 0.009 & (0.099) & 0.014 & (0.118)\\
$1_{SwC}1_{W}\lambda$ & 0.039*** & (0.008) & 0.039*** & (0.008)\\
$\beta_{M}$ & -0.023 & (0.047) & -0.028 & (0.048)\\
$\beta_{F_{L}}$ & -0.002 & (0.005) & -0.002 & (0.004)\\
$\beta_{t_{min}}$ & 2.488 & (3.265) & 3.071 & (4.022)\\
$-s_{N}1_{B}$ & -13.443 & (18.509) & -8.605 & (19.412)\\
$-s_{P}$ & -10.689*** & (4.020) & -11.024** & (4.601)\\
$1_{SwC}g_{N}$ & -2.385 & (11.311) & 14.282 & (16.810)\\
$1_{SwC}g_{P}$ & 11.011*** & (3.935) & 11.693** & (4.728)\\
$\beta_{AP}$ & 27.160** & (13.379) & 28.119** & (15.186)\\
$\beta_{APD}$ & 15.297 & (9.316) & 14.160 & (10.081)\\
$1_{N}$ & 13.866 & (9.724) & 3.115 & (6.256)\\
$Intercept$ & 31.643*** & (8.143) & 37.805*** & (8.936)\\
Year FX & \multicolumn{2}{c}{Yes} & \multicolumn{2}{c}{Yes}\\
Firm FX & \multicolumn{2}{c}{Yes} & \multicolumn{2}{c}{Yes}\\
Within $R^{2}$ & \multicolumn{2}{c}{0.087} & \multicolumn{2}{c}{0.099}\\
Between $R^{2}$ & \multicolumn{2}{c}{0.636} & \multicolumn{2}{c}{0.662}\\
Overall $R^{2}$ & \multicolumn{2}{c}{0.487} & \multicolumn{2}{c}{0.520}\\
Observations & \multicolumn{2}{c}{1,279} & \multicolumn{2}{c}{1,202}%
\\\hline\hline
\end{tabular}
\\
&  \text{(a) The dependent variable is }\frac{Bribe}{workers}\text{ in column
1 and }\frac{Rep\_Bribe}{workers}\text{ in column 2;}\\
&  \text{(b) Robust standard errors in parentheses are clustered by firm
id's;}\\
&  \text{(c) }\Pi_{N}\text{, }\Delta\Pi\text{, and }M\text{ are measured in
thousand US \$;}\\
&  \text{(d) ***, **, * - statistical significance at 1\%, 5\%, 10\% level.}%
\end{align*}
}

\normalsize

\begin{align*}
&  \text{Table B3. Baseline specification with the legal status of firms.}\\
&
\begin{tabular}{cll}
\hline\hline
Coefficients & \multicolumn{2}{c}{$\frac{Bribe}{workers}$} \\ \hline
$\beta _{\Pi _{N}}(1-1_{legal})$ & -1.381  & (1.414) \\ 
$\beta _{\Pi _{N}}1_{legal}$ & 6.539***  & (2.473) \\ 
$1_{NS}(1-1_{legal})$ & -0.064 & (0.068) \\ 
$1_{NS}1_{legal}$ & -0.071  & (0.134) \\ 
$1_{SwC}1_{W}\lambda (1-1_{legal})$ & 0.046*** & (0.009) \\ 
$1_{SwC}1_{W}\lambda 1_{legal}$ & 0.187  & (0.122) \\ 
$\beta _{M}$ & -0.021 & (0.048) \\ 
$\beta _{F_{L}}$ & -0.002 & (0.005) \\ 
$\beta _{t_{min}}$ & 1.064 & (3.278) \\ 
$-s_{N}1_{B}(1-1_{legal})$ & 29.900 & (53.809) \\ 
$-s_{N}1_{B}1_{legal}$ & -37.350  & (56.117) \\ 
$-s_{P}(1-1_{legal})$ & -6.099 & (4.063) \\ 
$-s_{P}1_{legal}$ & -5.206 & (7.289) \\ 
$1_{SwC}g(1-1_{legal})$ &  6.039 & (4.279) \\ 
$1_{SwC}g1_{legal}$ &  5.474 & (7.501) \\ 
$\beta _{AP}$ & 26.827** & (11.144) \\ 
$\beta _{APD}$ & 15.308 & (9.827) \\ 
$1_{legal}$ & -1.403  & (5.554) \\ 
$Intercept$ & 36.139*** & (7.973) \\ 
Year FX & \multicolumn{2}{c}{Yes} \\ 
Firm FX & \multicolumn{2}{c}{Yes} \\ 
Within $R^{2}$ & \multicolumn{2}{c}{0.090} \\ 
Between $R^{2}$ & \multicolumn{2}{c}{0.636} \\ 
Overall $R^{2}$ & \multicolumn{2}{c}{0.487} \\ 
Observations & \multicolumn{2}{c}{1,279} \\ \hline\hline
\end{tabular}
\end{align*}

\bigskip

\section*{C. Data collection and construction of variables}

\subsection*{C.1 Data collection}
The data contain responses of senior managers/accountants of 501 firms
operating in Tajikistan in 2012-2014. The collection of data was commissioned and funded by the Organization for Security and Co-operation in
Europe. The data were collected by Research Center SHARQ, Tajikistan, between
July 27, 2015 and September 18, 2015. Research Center SHARQ had prior
experience collecting firm-level data in Tajikistan for the World Bank's
Enterprise Surveys, Asian Development Bank's enterprise-level surveys, and
United Nation Development Programme's enterprise-level surveys.

The stratified random sampling procedure was used to construct the sample for
the study. The sample was stratified by geographical location and industry
type. The sample covers all administrative units (city of Dushanbe, Khatlon
province, Districts of Republican\ Subordination (DRS), Sughd province, and
Mountainous Badakhshon Autonomous Province (MBAP)) and all industries:
construction, manufacturing, trade, agriculture, and other services. The
sample of 2,000 firms was constructed based on the population of firms in
Tajikistan in 2015 (24,525 establishments), according to the data presented by
the State Statistical Committee of Tajikistan in June of 2015. Table C1
presents the population of firms used as the sample frame. Table C2 presents
the sample of 2,000 firms based on the sample frame, and Table C3 presents the
achieved sample of 501 firms. The sampling procedure was similar to those
adopted by the World Bank's Economic Surveys\footnote{See
\url{https://www.enterprisesurveys.org/en/methodology} for details.}.

Respondents had an option to answer the questionnaire either in Tajik or in
Russian. To improve accuracy of responses, respondents were assured that the
data were to be used anonymously and in aggregate form. To achieve higher
response rates, each respondent was guaranteed to receive US\$ 8 upon
completion of an interview, which was equivalent to 1.25 of an average daily
salary in September 2015. Table C4 describes variables used in the analysis in
the paper, and Table C5 lists codes and names of 65 tax districts.

The achieved sample and properties of the data are similar to those of the
World Bank's Enterprise Surveys. The World Bank's Business Environment and
Enterprise Performance Survey (BEEPS V) in Tajikistan in 2013-2014 is the
closest reference in terms of the time of data collection and sample size. The
BEEPS V study is a part of the World Bank's Enterprise Surveys, and it had
similar objective of collecting firm-level data on business practices and
administrative environment in Tajikistan in 2013-2014.

All firm-level studies of business practices and corruption, including ours,
have relatively low response rates. The response rate in our study was 25.05\%
(501 responses out of 2000 sampled). But, not all firms that agreed to
participate in the survey gave complete answers about variables of interest.
Excluding those firms that did not provide complete answers, the response rate
in our study was 21.45\% (429 complete responses out of 2000 sampled). This
response rate is similar to the one in the BEEPS V study in 2014 (our
reference study). The BEEPS V study had response rate of 22\% (359 complete
responses out of 1620 sampled). Because all such studies require firms'
disclosure of sensitive information (e.g. unreported revenues, unpaid taxes,
operation without permits), the selection of subjects is driven by the ability
of interviewers to convince subjects of confidentiality of disclosed data. For
this reason, one approach to eliminate the selection bias in these studies is
to control for interviewer fixed effects. Since we have panel data with firm
fixed effects and the same interviewer collected data from the same firm in
all years, the interviewer fixed effects are accounted for by the firm fixed effects.

The comparison of the achieved sample of 501 firms (Table C3) with the
constructed sample of 2000 firms (Table C2) shows that the achieved sample
over-represents firms in categories B (manufacturing) and D (wholesale and
retail trade). For example, the probability of a firm falling into category B
in the achieved sample is 25\% versus 10\% in the population of firms. The
probability of a firm falling into category D in the achieved sample is 33\%
versus 24\% in the population of firms.

Table C6 presents observable characteristics (reported profit, market value of
capital, number of workers, number of permits, official paperwork cost,
official paperwork processing time, and the number of bureaucratic disputes)
of firms reporting positive delays and aggregate bribes, reporting positive
delays and zero aggregate bribes, reporting neither delays nor bribes, and
those with missing delay or bribes data. If firms that agreed to participate
in the survey but did not disclose sensitive information about bribes and
delays are similar to the firms that decided not to participate in the survey,
we can compare sub-samples with missing and complete delay and bribes data to
get an idea about how firms that selected to participate in the survey are
different from those that chose not to do so. The last two columns in Table C6
give this information. The comparison of firms with complete bribes and delay
data and firms with missing bribes and delay data shows that reported profits
and official paperwork processing times are different, and the difference is
statistically significant at 1\% level. In addition, the difference in market
value of capital across the two types of firms is statistically significant at
10\% level. Differences in other variables are not statistically significant.
These differences suggest that the achieved sample (429 firms)
under-represents smaller firms with lower levels of capital and fewer
paperwork/certification requirements relatively to the firms in the sample
that chose not to participate in the survey.

\subsection*{C.2 Construction of variables}
We describe the construction of variables used in the estimation of specification EM1.1 in the main text. The summary statistics of these variables are reported in Table 3 in the main text.

\paragraph*{Firms' types ($1_{N}$ and $1_{P}$):} As we argue in the
paper, high-profit firms have an incentive to hide their profits to reduce
bribe demands from corrupt bureaucrats. Typically, such high-profit firms
over-state their operating costs and under-state revenues. To calculate the
size of hidden profit, we ask each firm to report its assessment of
under-reported revenue and over-stated cost of a similar firm in the same line
of business\footnote{The exact formulation of the question was, "We understand
that some establishments face barriers when complying with tax regulations.
What percentage of total annual sales/total annual costs do establishments of
the same size and operating in the same line of business report in their tax
reports?"
\par
It is the standard practice to elicit sensitive information such as this one
in the indirect form (see, for example, \cite{Svensson2003}). For instance, in the
World Bank's BEEPS V Enterprise Survey in Tajikistan in 2013, sensitive
information about firms' informal payments and compliance was elicited in the
same indirect form. When eliciting information about under-reported revenues
and over-stated costs, each firm was assured that the disclosed information
would not be reported to administrative authorities.}. We calculate adjusted
profit for each firm by adjusting the firm's reported revenue for its own
assessment of under-reported revenue and reported cost for its own assessment
of over-stated cost. If we use the adjusted profit to evaluate the
distribution of firms by profitability, we find that about 20\% of firms have
negative profit, 10\% have zero profit, and 70\% have positive profit (see
Table 1 in the main text).

We use adjusted profits to separate firms into two groups. Firms with positive
adjusted profits represent $P$-type firms and are indicated by $1_{P}$, while
firms with zero or negative adjusted profits represent $N$-type firms and are
indicated by $1_{N}=1_{P}-1$. The median bribe of firms with zero or negative
profits in the data is zero, which is consistent with the theoretical assumption that the
$N$-type firm does not participate in the bribe contest and does not pay
extortionary or non-extortionary bribes. Note that adjusted profits of some
$N$-type firms, while negative, still exceed what they report in tax reports.
This is because some firms with negative or zero adjusted profits have
positive differences between adjusted and reported profits.

\paragraph*{Profit of the $N$-type firm ($\Pi_{N}$):} We use
profit that each firm reports in its tax report as a measure for $\Pi_{N}$.
Specifically, we take the difference between annual total sales and annual
total costs reported to the tax service by each firm in each year. As Table 3 in the main text
shows, the mean reported profit is US \$ -56 thousand, and the overwhelming
majority of the firms (about 92\%) report negative or zero profit. The use of
the reported profit to measure $\Pi_{N}$ is consistent with the assumption of perfect
observability of $\Pi_{N}$ in the model of corruption.

\paragraph*{Positive profit increment ($\Delta\Pi(\Gamma)$):} We use the
difference between adjusted and reported profits as a measure for the positive
profit increment $\Delta\Pi(\Gamma)$. Whether corrupt bureaucrats observe the
unreported difference in profits is an important question that we attempt to
answer in the empirical exercise. 
The mean of the positive profit increment is US \$ 423 thousand, with the minimum of US \$ 0 and the maximum of US \$ 47,734 thousand. 

\paragraph*{Bureaucratic burden ($All\_papers$):} Different firms have
different bureaucratic requirements. To measure differences in the
bureaucratic burden, we ask firms to report the total number of required legal
papers ($All\_papers$) that firms have to possess according to the
law\footnote{The exact formulation of the question was, "Do you possess all
legal papers, certificates, licenses, permits and resolutions from all
administrative and supervisory officials that are required to achieve a fully
legal status according to the law? Indicate the total number of required legal
papers and the share of legal papers in your possession."}. The required legal
papers include all certificates, permits, and licenses required for operation
in a fiscal year. The mean number of required legal papers is 8.1, the minimum
is 1 paper, and the maximum is 30 papers.

\paragraph*{Market value of capital ($M$):} We use firms' reported market
value of capital (equipment, machinery, land, and real estate) to measure $M$.
The market value of capital $M$ captures each firm's outside option, if the
firm decides to shut down and sell off its capital. In the data, the mean
market value of capital is US \$ 554 thousand, and the value varies between US
\$ 66 and US \$ 63,645 thousand.

\paragraph*{Official paperwork processing cost ($F_{L}$):} To measure
$F_{L}$, we use firms' reports of annual aggregate expenditures on completion
of all bureaucratic requirements, if these requirements were to be satisfied
strictly according to the law. This measure aggregates bureaucratic
expenditures only through official channels and includes tax payments,
certification and registration fees, and all associated compliance
costs\footnote{The exact formulation of the question was, \textquotedblleft
What would be the aggregate expenses only through official channels (including
notary fees and expenses on lawyers) on full compliance with tax requirements,
customs requirements, acquisition of licenses, certificates, registration
papers, and other administrative papers to ensure legal and transparent
operation of your firm according to the law?\textquotedblright\ }. The mean of
this measure is US \$ 40,981, with the minimum of US \$ 10.6 and the maximum
of US \$ 9,361 thousand.

\paragraph*{Aggregate bribe ($Bribe$):} We use two measures of aggregate
bribe. The first measure is constructed indirectly. We ask each firm to report
its actual aggregate expenditures on satisfying bureaucratic requirements
(through official and unofficial channels)\footnote{The exact formulation of
the question was, \textquotedblleft What were the actual expenses on
compliance with tax requirements, customs requirements, acquisition of
licenses, certificates, registration papers, and other administrative papers
to ensure operation of your firm?\textquotedblright}. Then, we subtract from
this reported value the firm's measure for official paperwork processing cost
(measure of $F_{L}$). The resulting difference is our indirect measure of
aggregate bribe ($Bribe$) that we use to estimate empirical specification EM1.1 of the main text. The main advantages of this indirect measure are
the higher measurement accuracy and the higher response rate of firms, as firms did not
have to explicitly admit to payment of bribes when reporting actual and
official bureaucratic expenditures. The mean of the indirect measure of
aggregate bribe is US \$ 1,994, the minimum is US \$ 0, and the maximum is US
\$ 800,000.

The second measure uses direct reports of annual aggregate bribes. Each firm
reported annual aggregate informal payment as a share of annual revenue made
by a firm of similar size in the same line of business ($Rep\_bribe$%
)\footnote{The exact formulation of the question was, "On average, what share
of your annual sales does an establishment of your size and in your line of
business spend on informal payments (including gifts and services to
government officials and involuntary charity contributions)?"}. This is the
same measure of bribes that is used in the WBES data, and we use this directly measured aggregate bribe to estimate the alternative empirical specification EM1.2 that serves as our robustness check (see the empirical specification EM1.1 of the main text and subsequent discussion). The mean reported
aggregate bribe is US \$ 2,195, with the minimum of US \$ 0 and the maximum of
US \$ 800,000. The mean difference between direct and indirect measures of
aggregate bribes is US \$ -0.4, and the 95\% confidence interval is
$[-13.9,13.2]$. The largest difference is in response rates. We have 494 (out
of 501) indirectly measured aggregate bribe reports and 449 directly measured
aggregate bribe reports.

\paragraph*{Bureaucratic disputes ($1_{disp}$):} Our measure of bribes combines
extortionary and non-extortionary bribes, as reports of bribes made to
individual bureaucrats are impossible to collect. To distinguish between
extortionary and non-extortionary bribes, we rely on the definitions in
\cite{Ayres1997} and \cite{KhalilLawareeYun2010}. Following \cite{Ayres1997},
we define extortionary bribery as the bribery initiated by bureaucrats. To
capture this type of bribery, we use information about disputes that are
initiated by bureaucrats rather than firms. We ask firms to report the total
number of disputes that government officials initiated by citing lack of
required legal papers or inconsistencies in available legal
papers\footnote{The exact formulation of the question was, "Report the total
number of cases when tax, customs, or any other administrative officials
initiated disputes because of missing tax or customs declarations,
certificates, or other required permits or inconsistencies in these papers."}.
The indicator for the positive number of bureaucratic disputes is our indicator of extortionary
bribery, and it takes the value of one ($1_{disp}=1$) for a firm reporting at
least one bureaucratic dispute and the value of zero ($1_{disp}=0$) for a firm
reporting no such disputes. More than 22\% of firms in the sample reported at
least one such dispute in fiscal year. The mean number of disputes among firms
with at least one dispute is 1.4, with the maximum of 10 disputes.

\paragraph*{Official paperwork processing time ($t_{min}$):} To measure
$t_{min}$, we use firms' reports of the total number of days necessary to
satisfy all bureaucratic requirements only through official channels and
according to the law\footnote{The exact formulation was, \textquotedblleft How
many days it would have taken to obtain all required papers only through
official channels (complete tax requirements, customs requirements, acquire
licenses, certificates, registration papers, and other administrative papers
to ensure operation of your firm) to achieve the fully legal status according
to the law?\textquotedblright\ }. The mean number of days is 11.6 days, with
the minimum of 0 days and the maximum of 150 days. Note that although there
are time limits for paperwork processing times, these time limits are often exceeded.

\paragraph*{Bureaucratic delay ($m_{N},m_{P},n$):} There are three types
of time periods that firms spend with bureaucrats: bureaucratic delay of the
$N$-type firm due to extortionary bribery ($m_{N}$), bureaucratic delay of the
$P$-type firm due to extortionary bribery ($m_{P}$), and the $P$-type firm's time in the bribe contest ($n$). Just as with separate bribe reports,
we would like to have separate measures for these time periods. Unfortunately, such measures are
unavailable, because of the complexity of informal transfers and the general inability of firms to classify bribery and associated time losses into different categories.

To construct measures for these
time periods, we first construct a measure of unspecified bureaucratic delay
($Delay$). To do so, we subtract the measure of official paperwork processing
time ($t_{min}$)\ from the actual number of days that firms spend with
bureaucrats while satisfying bureaucratic requirements in fiscal
year\footnote{The exact formulation of the question was, \textquotedblleft How
many days did it take you to complete tax requirements, customs requirements,
acquire licenses, certificates, registration papers, and other administrative
papers to ensure operation of your firm?\textquotedblright\ }. The resulting
difference is our measure of unspecified bureaucratic delay. This measure
captures the total time that firms spend with bureaucrats in excess of what is
permissible by law, and it includes the negotiation time with bureaucrats when
trying to gain access to government resources (time in bribe contests) and the
intentionally created bureaucratic delay when corrupt bureaucrats forcibly
extract bribes (bureaucratic delay). The average unspecified bureaucratic
delay is 4.5 days, the minimum is 0 days, and the maximum is 100 days.

To measure $m_{P}$, we interact the unspecified bureaucratic delay ($Delay$)
with the indicator of the $P$-type firm ($1_{P}$) and the indicator of
bureaucratic disputes ($1_{disp}$), so that $m_{P}=Delay1_{P}1_{disp}$. To
measure $m_{N}$, we interact the unspecified bureaucratic delay with the
indicator of the $N$-type firm ($1_N=1-1_{P}$) and the indicator of
bureaucratic disputes, so that $m_{N}=Delay(1-1_{P})1_{disp}$. To measure $n$,
we interact the unspecified bureaucratic delay with the indicator of the
$P$-type firm, so that $n=$ $Delay1_{P}$.

\begin{align*}
&  \text{Table C1. Population of firms (Source: State Statistical Committee of
Tajikistan, 2015).}\\
&
\begin{tabular}
[c]{ccccccc}\hline\hline
{\small Industrial } & {\small Dushanbe } & {\small Sughd } & {\small Khatlon
} & {\small DRS } & {\small MBAP } & {\small Total}\\
{\small category} & {\small city} & {\small Province} & {\small Province} &
({\small Districts} & ({\small Mount.} & \\
&  &  &  & {\small of Repub.} & {\small Badakhshon} & \\
&  &  &  & {\small Subord.)} & {\small Autonomous} & \\
&  &  &  &  & {\small \ Province)} & \\\hline
{\small A. Agriculture; } &  &  &  &  &  & \\
{\small Forestry \& fishing;} & {\small 34} & {\small 2131} & {\small 5500} &
{\small 505} & {\small 111} & {\small 8281}\\
{\small B. Manufacturing; Mining \& } &  &  &  &  &  & \\
{\small quarrying; Electricity, gas,} &  &  &  &  &  & \\
{\small steam, air \& water supply;} & {\small 750} & {\small 892} &
{\small 579} & {\small 212} & {\small 54} & {\small 2487}\\
{\small C. Construction; Real estate,} &  &  &  &  &  & \\
{\small professional, scientific, } &  &  &  &  &  & \\
{\small technical, administrative \& } &  &  &  &  &  & \\
{\small support service activities;} & {\small 2271} & {\small 893} &
{\small 810} & {\small 285} & {\small 171} & {\small 4430}\\
{\small D. Wholesale \& retail trade; } &  &  &  &  &  & \\
{\small Repair of motor vehicles \&} &  &  &  &  &  & \\
{\small motorcycles; Accom- } &  &  &  &  &  & \\
{\small modation \& food services;} &  &  &  &  &  & \\
{\small Arts, entertainment, } &  &  &  &  &  & \\
{\small recreation \& other services;} & {\small 2415} & {\small 2001} &
{\small 1375} & {\small 636} & {\small 222} & {\small 6649}\\
{\small E. Transportation, storage,} &  &  &  &  &  & \\
{\small information \& communication;} & {\small 548} & {\small 272} &
{\small 126} & {\small 37} & {\small 55} & {\small 1038}\\
{\small F. Financial \& insurance } &  &  &  &  &  & \\
{\small activities;} & {\small 165} & {\small 55} & {\small 27} & {\small 6} &
{\small 8} & {\small 261}\\
{\small G. Human health, social } &  &  &  &  &  & \\
{\small work \& education;} & {\small 410} & {\small 416} & {\small 364} &
{\small 104} & {\small 85} & {\small 1379}\\
{\small Total} & {\small 6593} & {\small 6660} & {\small 8781} & {\small 1785}
& {\small 706} & {\small 24525}\\\hline\hline
\end{tabular}
\end{align*}

\begin{align*}
&  \text{Table C2. The constructed sample based on the population of firms
(2000 firms).}\\
&
\begin{tabular}
[c]{ccccccc}\hline\hline
{\small Industrial } & {\small Dushanbe } & {\small Sughd } & {\small Khatlon
} & {\small DRS } & {\small MBAP } & {\small Total}\\
{\small category} & {\small city} & {\small Province} & {\small Province} &
({\small Districts} & ({\small Mount.} & \\
&  &  &  & {\small of Repub.} & {\small Badakhshon} & \\
&  &  &  & {\small Subord.)} & {\small Autonomous} & \\
&  &  &  &  & {\small \ Province)} & \\\hline
{\small A. Agriculture; } &  &  &  &  &  & \\
{\small Forestry \& fishing;} & {\small 4} & {\small 172} & {\small 448} &
{\small 40} & {\small 8} & {\small 672}\\
{\small B. Manufacturing; Mining \& } &  &  &  &  &  & \\
{\small quarrying; Electricity, gas,} &  &  &  &  &  & \\
{\small steam, air \& water supply;} & {\small 60} & {\small 72} & {\small 48}
& {\small 16} & {\small 4} & {\small 200}\\
{\small C. Construction; Real estate,} &  &  &  &  &  & \\
{\small professional, scientific, } &  &  &  &  &  & \\
{\small technical, administrative \& } &  &  &  &  &  & \\
{\small support service activities;} & {\small 184} & {\small 72} &
{\small 68} & {\small 24} & {\small 12} & {\small 360}\\
{\small D. Wholesale \& retail trade; } &  &  &  &  &  & \\
{\small Repair of motor vehicles \&} &  &  &  &  &  & \\
{\small motorcycles; Accom- } &  &  &  &  &  & \\
{\small modation \& food services;} &  &  &  &  &  & \\
{\small Arts, entertainment, } &  &  &  &  &  & \\
{\small recreation \& other services;} & {\small 200} & {\small 164} &
{\small 112} & {\small 52} & {\small 20} & {\small 548}\\
{\small E. Transportation, storage,} &  &  &  &  &  & \\
{\small information \& communication;} & {\small 44} & {\small 24} &
{\small 12} & {\small 4} & {\small 4} & {\small 88}\\
{\small F. Financial \& insurance } &  &  &  &  &  & \\
{\small activities;} & {\small 12} & {\small 4} & {\small 4} & {\small 0} &
{\small 0} & {\small 20}\\
{\small G. Human health, social } &  &  &  &  &  & \\
{\small work \& education;} & {\small 32} & {\small 36} & {\small 32} &
{\small 8} & {\small 4} & {\small 112}\\
{\small Total} & {\small 536} & {\small 544} & {\small 724} & {\small 144} &
{\small 52} & {\small 2000}\\\hline\hline
\end{tabular}
\end{align*}
\bigskip

\qquad\qquad%
\begin{align*}
&  \text{Table C3. The achieved sample (501 firms).}\\
&
\begin{tabular}
[c]{ccccccc}\hline\hline
{\small Industrial } & {\small Dushanbe } & {\small Sughd } & {\small Khatlon
} & {\small DRS } & {\small MBAP } & {\small Total}\\
{\small category} & {\small city} & {\small Province} & {\small Province} &
({\small Districts} & ({\small Mount.} & \\
&  &  &  & {\small of Repub.} & {\small Badakhshon} & \\
&  &  &  & {\small Subord.)} & {\small Autonomous} & \\
&  &  &  &  & {\small \ Province)} & \\\hline
{\small A. Agriculture; } &  &  &  &  &  & \\
{\small Forestry \& fishing;} & {\small 0} & {\small 25} & {\small 85} &
{\small 9} & {\small 2} & {\small 121}\\
{\small B. Manufacturing; Mining \& } &  &  &  &  &  & \\
{\small quarrying; Electricity, gas,} &  &  &  &  &  & \\
{\small steam, air \& water supply;} & {\small 36} & {\small 49} & {\small 27}
& {\small 10} & {\small 2} & {\small 124}\\
{\small C. Construction; Real estate,} &  &  &  &  &  & \\
{\small professional, scientific, } &  &  &  &  &  & \\
{\small technical, administrative \& } &  &  &  &  &  & \\
{\small support service activities;} & {\small 14} & {\small 14} & {\small 12}
& {\small 10} & {\small 3} & {\small 53}\\
{\small D. Wholesale \& retail trade; } &  &  &  &  &  & \\
{\small Repair of motor vehicles \&} &  &  &  &  &  & \\
{\small motorcycles; Accom- } &  &  &  &  &  & \\
{\small modation \& food services;} &  &  &  &  &  & \\
{\small Arts, entertainment, } &  &  &  &  &  & \\
{\small recreation \& other services;} & {\small 52} & {\small 40} &
{\small 42} & {\small 26} & {\small 6} & {\small 166}\\
{\small E. Transportation, storage,} &  &  &  &  &  & \\
{\small information \& communication;} & {\small 8} & {\small 13} & {\small 8}
& {\small 1} & {\small 0} & {\small 30}\\
{\small F. Financial \& insurance } &  &  &  &  &  & \\
{\small activities;} & {\small 0} & {\small 1} & {\small 0} & {\small 0} &
{\small 0} & {\small 1}\\
{\small G. Human health, social } &  &  &  &  &  & \\
{\small work \& education;} & {\small 4} & {\small 2} & {\small 0} &
{\small 0} & {\small 0} & {\small 6}\\
{\small Total} & {\small 114} & {\small 144} & {\small 174} & {\small 56} &
{\small 13} & {\small 501}\\\hline\hline
\end{tabular}
\end{align*}

\begin{align*}
&  \text{Table C4. Description of variables.}\\
&
\begin{tabular}
[c]{cc}\hline\hline
{\small Variable name} & {\small Description}\\\hline
\multicolumn{1}{l}{{\small id}} & \multicolumn{1}{l}{{\small Firm
identification number}}\\
\multicolumn{1}{l}{{\small year}} & \multicolumn{1}{l}{{\small Year of
reported data}}\\
\multicolumn{1}{l}{{\small location}} & \multicolumn{1}{l}{{\small Region: "1"
- Dushanbe city, "2" - Khatlon province, "3" - DRS,}}\\
\multicolumn{1}{l}{} & \multicolumn{1}{l}{{\small "4" - Sughd province, and
"5" - MBAP}}\\
\multicolumn{1}{l}{{\small tax\_district}} & \multicolumn{1}{l}{{\small Code
number for tax district}}\\
\multicolumn{1}{l}{{\small year\_estab}} & \multicolumn{1}{l}{{\small Year,
when the firm was established}}\\
\multicolumn{1}{l}{{\small ownership}} & \multicolumn{1}{l}{{\small Ownership
status: "1" - sole proprietorship, "2" - joint stock}}\\
\multicolumn{1}{l}{} & \multicolumn{1}{l}{{\small company, "3" - cooperative,
"4" - corporation, "5" - other privately}}\\
\multicolumn{1}{l}{} & \multicolumn{1}{l}{{\small owned company, "6" - other
ownership form}}\\
\multicolumn{1}{l}{{\small state\_share}} & \multicolumn{1}{l}{{\small Share
of state ownership (in percent)}}\\
\multicolumn{1}{l}{{\small industry\_group}} &
\multicolumn{1}{l}{{\small Industrial category: ISIC rev. 4 two-digit
industrial division code}}\\
\multicolumn{1}{l}{{\small industry}} & \multicolumn{1}{l}{{\small Industrial
sub-category: ISIC rev. 4 four-digit industrial class code}}\\
{\small sector\_outputTJ} & \multicolumn{1}{l}{{\small Total output in TJS in
agriculture, construction, manufacturing, trade,}}\\
& \multicolumn{1}{l}{{\small transportation and communication, services
(source: State Stat. Comm.)}}\\
\multicolumn{1}{l}{{\small revenueTJ}} & \multicolumn{1}{l}{{\small Total
sales reported to the tax service in TJ somonis}}\\
\multicolumn{1}{l}{{\small costTJ}} & \multicolumn{1}{l}{{\small Total
production cost reported to the tax service in TJ somonis}}\\
\multicolumn{1}{l}{{\small capital\_costTJ}} &
\multicolumn{1}{l}{{\small Market value of capital in TJ somonis}}\\
\multicolumn{1}{l}{{\small reported\_sales}} &
\multicolumn{1}{l}{{\small Share of actual sales reported to the tax service
(in percent of actual sales)}}\\
\multicolumn{1}{l}{{\small reported\_cost}} &
\multicolumn{1}{l}{{\small Excess production cost reported to the tax service
(in percent of actual cost)}}\\
\multicolumn{1}{l}{{\small official\_feeTJ}} &
\multicolumn{1}{l}{{\small Official paperwork cost in TJ somonis}}\\
\multicolumn{1}{l}{{\small official\_days}} &
\multicolumn{1}{l}{{\small Official paperwork time in days}}\\
\multicolumn{1}{l}{{\small actual\_feeTJ}} & \multicolumn{1}{l}{{\small Actual
expenditures on completing paperwork in TJ somonis}}\\
\multicolumn{1}{l}{{\small actual\_days}} & \multicolumn{1}{l}{{\small Actual
time spent on completing paperwork in days}}\\
\multicolumn{1}{l}{{\small allworkers}} & \multicolumn{1}{l}{{\small Number of
employees}}\\
\multicolumn{1}{l}{{\small paper\_number}} & \multicolumn{1}{l}{{\small Total
number of required legal papers (certificates, permits, licenses etc.)}}\\
\multicolumn{1}{l}{{\small rep\_bribes}} & \multicolumn{1}{l}{{\small Reported
annual aggregate bribe in TJ somonis}}\\
\multicolumn{1}{l}{{\small dispute\_number}} &
\multicolumn{1}{l}{{\small Annual number of disputes }}\\
\multicolumn{1}{l}{{\small interviewer}} &
\multicolumn{1}{l}{{\small Interviewer code}}\\
\multicolumn{1}{l}{{\small date}} & \multicolumn{1}{l}{{\small Date of the
interview}}\\\hline\hline
\end{tabular}
\end{align*}

\begin{align*}
&  \text{Table C5. Tax district names and codes.}\\
&
\begin{tabular}
[c]{cccccccc}\hline\hline
{\small District } & {\small District} &  & {\small District } &
{\small District} &  & {\small District } & {\small District}\\
{\small name} & {\small code} &  & {\small name} & {\small code} &  &
{\small name} & {\small code}\\\hline
{\small Asht} & {\small 1} &  & {\small Kuhistoni } &  &  & {\small Spitamen}
& {\small 51}\\
&  &  & {\small Mastchohi} & {\small 26} &  & {\small Tavildara} &
{\small 52}\\
{\small Ayni} & {\small 2} &  & {\small Kulob} & {\small 27} &  &
{\small Temurmalik} & {\small 53}\\
{\small Baljuvon} & {\small 3} &  & {\small Mastchoh} & {\small 28} &  &
{\small Tojikobod} & {\small 54}\\
{\small Bobojon Gafurov} & {\small 4} &  & {\small Muminobod} & {\small 29} &
& {\small Tursunzoda} & {\small 55}\\
{\small Bohtar} & {\small 5} &  & {\small Murgob} & {\small 30} &  &
{\small Vahdat} & {\small 56}\\
{\small Danghara} & {\small 6} &  & {\small Norak} & {\small 31} &  &
{\small Vahsh} & {\small 57}\\
{\small Darvoz} & {\small 7} &  & {\small Nosiri Hisrav} & {\small 32} &  &
{\small Vanj} & {\small 58}\\
{\small Faizobod} & {\small 8} &  & {\small Nurobod} & {\small 33} &  &
{\small Varzob} & {\small 59}\\
{\small Farhor} & {\small 9} &  & {\small Panj} & {\small 34} &  &
{\small Vose'} & {\small 60}\\
{\small Firdavsi} & {\small 10} &  & {\small Panjakent} & {\small 35} &  &
{\small Yovon} & {\small 61}\\
{\small Gonchi} & {\small 11} &  & {\small Kabodiyon} & {\small 36} &  &
{\small Zafarobod} & {\small 62}\\
{\small Hamadoni} & {\small 12} &  & {\small Kumsangir} & {\small 37} &  &
{\small Kurgon Teppa} & {\small 63}\\
{\small Hisor} & {\small 13} &  & {\small Rasht} & {\small 38} &  &
{\small Khujand} & {\small 64}\\
{\small Ibn Sino} & {\small 14} &  & {\small Rogun} & {\small 39} &  &
{\small Khorog} & {\small 65}\\
{\small Isfara} & {\small 15} &  & {\small Roshtqal'a} & {\small 40} &  &  &
\\
{\small Ishkoshim} & {\small 16} &  & {\small Rudaki} & {\small 41} &  &  & \\
{\small Ismoili Somoni} & {\small 17} &  & {\small Rumi} & {\small 42} &  &  &
\\
{\small Istaravshan} & {\small 18} &  & {\small Rushon} & {\small 43} &  &  &
\\
{\small Jabbor Rasulov} & {\small 19} &  & {\small Sarband} & {\small 44} &  &
& \\
{\small Jilikul} & {\small 20} &  & {\small Shahrinav} & {\small 45} &  &  &
\\
{\small Jirgatol} & {\small 21} &  & {\small Shahriston} & {\small 46} &  &  &
\\
{\small Jomi} & {\small 22} &  & {\small Shahritus} & {\small 47} &  &  & \\
{\small Khovaling} & {\small 23} &  & {\small Shohmansur} & {\small 48} &  &
& \\
{\small Khuroson} & {\small 24} &  & {\small Shugnon} & {\small 49} &  &  & \\
{\small Konibodom} & {\small 25} &  & {\small Shurobod} & {\small 50} &  &  &
\\\hline\hline
\end{tabular}
\end{align*}

\begin{align*}
&  \text{Table C6. Comparisons of firm characteristics (means in 2014 USD).}\\
&
\begin{tabular}
[c]{cccccc}\hline\hline
Variables & Delay \& & Delay \& & No delay \& & Complete bribe & Missing
delay\\
& bribes & no bribes & no bribes & \ \& delay data & or bribes data\\\hline
Reported profit & \multicolumn{1}{r}{\ -72,903} & \multicolumn{1}{r}{-6,947} &
\multicolumn{1}{r}{-23,372} & \multicolumn{1}{r}{-56,754} &
\ \ \ \ \ -112***\\
Market value of capital & \multicolumn{1}{r}{672,668} &
\multicolumn{1}{r}{384,854} & \multicolumn{1}{r}{\ 271,287} &
\multicolumn{1}{r}{560,690} & 91,162*\\
Workers (number) & \multicolumn{1}{r}{42.975} & \multicolumn{1}{r}{42.337} &
\multicolumn{1}{r}{39.440} & \multicolumn{1}{r}{42.102} & 90.000\\
Legal papers (number) & \multicolumn{1}{r}{7.988} & \multicolumn{1}{r}{8.196}
& \multicolumn{1}{r}{\ 8.623} & \multicolumn{1}{r}{8.158} & 6.850\\
Official paper. cost & \multicolumn{1}{r}{25,654} & \multicolumn{1}{r}{21,547}
& \multicolumn{1}{r}{89,919} & \multicolumn{1}{r}{40,981} & N.A.\\
Official paper. time & \multicolumn{1}{r}{11.476} & \multicolumn{1}{r}{8.272}
& \multicolumn{1}{r}{12.801} & \multicolumn{1}{r}{11.605} & \ \ \ 9.321***\\
Disputes & \multicolumn{1}{r}{0.243} & \multicolumn{1}{r}{0.304} &
\multicolumn{1}{r}{0.144} & \multicolumn{1}{r}{0.221} & 0.150\\
Number of firms & \multicolumn{1}{r}{1022} & \multicolumn{1}{r}{92} &
\multicolumn{1}{r}{361} & \multicolumn{1}{r}{1483} & \ \ \ 20\\\hline\hline
\multicolumn{6}{l}{{\small Note: ***, **, * - the sub-sample with missing data
is statistically different from the}}\\
\multicolumn{6}{l}{{\small sub-sample with complete data at 1\%, 5\%, 10\%
level\ in the Kruskal-Wallis test.}}%
\end{tabular}
\end{align*}

\begin{align*}
&  \text{Table C7. Summary statistics for normalized variables (all years in
2014 US \$).}\\
&
\begin{tabular}
[c]{cccc}\hline\hline
Variable name & Mean & Std. Err. & 95\% Confidence Interval\\\hline
\multicolumn{1}{l}{Reported profit per worker $\frac{\Pi_{{\small N}}%
}{workers}$ \ (thousand \$)} & -1.372 & 0.284 & [-1.929, -0.814]\\
\multicolumn{1}{l}{Unreported profit per worker $\frac{\Delta\Pi(\Gamma
)}{workers}$ (thousand \$)} & 11.105 & 3.299 & [4.632, 17.577]\\
\multicolumn{1}{l}{Required papers per worker $\frac{All\_papers}{workers}$
(number)} & 0.627 & 0.016 & [0.595, 0.658]\\
\multicolumn{1}{l}{Capital per worker $\frac{M}{workers}$ (thousand \$)} &
18.288 & 5.849 & [6.814, 29.763]\\
\multicolumn{1}{l}{Official process. cost per worker $\frac{F_{{\small L}}%
}{workers}$ (\$)} & 1338.4 & 334.7 & [681.6, 1995.3]\\
\multicolumn{1}{l}{Actual process. cost per worker $\frac{F}{workers}$ (\$)} &
1394.8 & 335.1 & [737.4, 2052.3]\\
\multicolumn{1}{l}{Aggregate bribe per worker $\frac{Bribe}{workers}$ (\$)} &
56.396 & 9.619 & [37.523, 75.268]\\
\multicolumn{1}{l}{Reported aggregate bribe per worker $\frac{Rep\_bribe}%
{workers}$ (\$)} & 57.006 & 9.638 & [38.095, 75.916]\\
\multicolumn{1}{l}{Official process. time per worker $\frac{t_{{\small min}}%
}{workers}$ (days)} & 0.818 & 0.030 & [0.758, 0.877]\\
\multicolumn{1}{l}{Actual process. time per worker $\frac{t}{workers}$ (days)}
& 1.185 & 0.047 & [1.094, 1.277]\\
\multicolumn{1}{l}{Bureaucratic delay per worker $\frac{Delay}{workers}$
(days)} & 0.367 & 0.022 & [0.324, 0.411]\\\hline\hline
\end{tabular}
\end{align*}

\end{document}